\documentclass[amsmath,floatfix,twocolumn,superscriptaddress]{revtex4-1}
\usepackage{amssymb,amsmath}
\usepackage{graphicx,array}
\usepackage{dcolumn}
\usepackage{psfrag}
\usepackage{bm}
\usepackage{color}
\usepackage{epstopdf}

\begin{document}

\title{Colossal reversible barocaloric effects in a plastic crystal mediated by lattice vibrations and ion diffusion}

\author{Ming Zeng}
\affiliation{Grup de Caracterizació de Materials, Departament de Física, EEBE and Barcelona Research Center in Multiscale Science and Engineering Universitat Politècnica de Catalunya, Av. Eduard Maristany 10-14, Barcelona 08019, Catalonia, Spain}

\author{Carlos Escorihuela-Sayalero}
\affiliation{Grup de Caracterizació de Materials, Departament de Física, EEBE and Barcelona Research Center in Multiscale Science and Engineering Universitat Politècnica de Catalunya, Av. Eduard Maristany 10-14, Barcelona 08019, Catalonia, Spain}

\author{Tamio Ikeshoji}
\affiliation{Mathematics for Advanced Materials Open Innovation Laboratory (MathAM-OIL), National Institute of Advanced Industrial Science and Technology (AIST), c/o Advanced Institute for Materials Research (AIMR), Tohoku University, Sendai 980-8577, Japan}

\author{Shigeyuki Takagi}
\affiliation{Institute for Materials Research (IMR), Tohoku University, Sendai 980-8577, Japan}

\author{Sangryun Kim}
\affiliation{Graduate School of Energy Convergence, Gwangju Institute of Science and Technology (GIST), 123 Cheomdangwagi-ro, Buk-gu, Gwangju 61005, Republic of Korea}

\author{Shin-ichi Orimo}
\affiliation{Advanced Institute for Materials Research (AIMR), Tohoku University, Sendai 980-8577, Japan}
\affiliation{Institute for Materials Research (IMR), Tohoku University, Sendai 980-8577, Japan}

\author{María Barrio}
\affiliation{Grup de Caracterizació de Materials, Departament de Física, EEBE and Barcelona Research Center in Multiscale Science and Engineering Universitat Politècnica de Catalunya, Av. Eduard Maristany 10-14, Barcelona 08019, Catalonia, Spain}

\author{Josep-Lluís Tamarit}
\affiliation{Grup de Caracterizació de Materials, Departament de Física, EEBE and Barcelona Research Center in Multiscale Science and Engineering Universitat Politècnica de Catalunya, Av. Eduard Maristany 10-14, Barcelona 08019, Catalonia, Spain}

\author{Pol Lloveras}
\email{pol.lloveras@upc.edu}
\affiliation{Grup de Caracterizació de Materials, Departament de Física, EEBE and Barcelona Research Center in Multiscale Science and Engineering Universitat Politècnica de Catalunya, Av. Eduard Maristany 10-14, Barcelona 08019, Catalonia, Spain}

\author{Claudio Cazorla}
\email{claudio.cazorla@upc.edu}
\affiliation{Grup de Caracterizació de Materials, Departament de Física, EEBE and Barcelona Research Center in Multiscale Science and Engineering Universitat Politècnica de Catalunya, Av. Eduard Maristany 10-14, Barcelona 08019, Catalonia, Spain}

\author{Kartik Sau}
\email{kartik.sau@gmail.com}
\affiliation{Advanced Institute for Materials Research (AIMR), Tohoku University, Sendai 980-8577, Japan}
\affiliation{Mathematics for Advanced Materials Open Innovation Laboratory (MathAM-OIL), National Institute of Advanced Industrial Science and Technology (AIST), c/o Advanced Institute for Materials Research (AIMR), Tohoku University, Sendai 980-8577, Japan}

\begin{abstract}
Solid-state methods for cooling and heating promise a more sustainable alternative to current  compression cycles of greenhouse gases and inefficient fuel-burning heaters. Barocaloric effects (BCE) driven by hydrostatic pressure ($p$) are especially encouraging in terms of large adiabatic temperature changes ($|\Delta T| \sim 10$~K) and colossal isothermal entropy changes ($|\Delta S| \sim 100$~J~K$^{-1}$~kg$^{-1}$). However, BCE typically require large pressure shifts due to irreversibility issues, and sizeable $|\Delta T|$ and $|\Delta S|$ seldom are realized in a same material. Here, we demonstrate the existence of colossal and reversible BCE in LiCB$_{11}$H$_{12}$, a well-known solid electrolyte, near its order-disorder phase transition at $\approx 380$~K. Specifically, for $\Delta p \approx 0.23$~(0.10)~GPa we measured $|\Delta S_{\rm rev}| = 280$~(200)~J~K$^{-1}$~kg$^{-1}$ and $|\Delta T_{\rm rev}| = 32$~(10)~K, which individually rival with state-of-the-art barocaloric shifts obtained under similar pressure conditions. Furthermore, over a wide temperature range, pressure shifts of the order of $0.1$~GPa yield huge reversible barocaloric strengths of $\approx 2$~J~K$^{-1}$~kg$^{-1}$~MPa$^{-1}$. Molecular dynamics simulations were carried out to quantify the role of lattice vibrations, molecular reorientations and ion diffusion on the disclosed colossal BCE. Interestingly, lattice vibrations were found to contribute the most to $|\Delta S|$ while the diffusion of lithium ions, despite adding up only slightly to the accompanying entropy change, was crucial in enabling the molecular order-disorder phase transition. Our work expands the knowledge on plastic crystals and should motivate the investigation of BCE in a variety of solid electrolytes displaying ion diffusion and concomitant molecular orientational disorder.   
\end{abstract}

\keywords{solid-state refrigeration, barocaloric effects, orientational order-disorder phase transition, lithium diffusion, molecular dynamics simulations}

\maketitle

\section{Introduction}
\label{sec:intro}
Solid-state methods for cooling and heating are energy efficient and ecologically friendly techniques with potential for solving the environmental problems posed by conventional refrigeration and heat pump technologies relying on compression cycles of greenhouse gases and inefficient traditional fuel-burning heaters \cite{Moya2020}. Under moderate magnetic, electric or mechanical field variations, auspicious caloric materials experience large adiabatic temperature variations ($|\Delta T| \sim 1$--$10$~K) as a result of phase transformations entailing large isothermal entropy changes ($|\Delta S| \sim 10$--$100$~J~K$^{-1}$~kg$^{-1}$) \cite{manosa13,Hou2022}. Solid-state cooling and heat pumping capitalize on such caloric effects for engineering refrigeration and heating cycles. From a practical point of view, large and reversible $|\Delta T|$ and $|\Delta S|$ are both necessary for achieving rapid and efficient devices under recursive application and removal of the driving fields. In terms of largest $|\Delta T|$ and $|\Delta S|$, mechanocaloric effects induced by uniaxial stress (elastocaloric effects) and hydrostatic pressure (barocaloric effects --BCE--) are among the most promising \cite{manosa17,cazorla19,lloveras21}.

Recently, colossal and reversible BCE ($|\Delta S_{\rm rev}| \ge 100$~J~K$^{-1}$~kg$^{-1}$) have been measured in several families of materials displaying order-disorder phase transitions under pressure shifts of the order of $0.1$~GPa \cite{lloveras19,li19,aznar20,aznar21,li22,imamura20,lloveras21b,mason22,Salvatori2022}. On one hand, there are plastic crystals like neopentane derivatives \cite{lloveras19,li19,aznar20}, adamantane derivatives \cite{aznar21,Salvatori2022} and carboranes \cite{li22} in which the underlying phase transitions involve molecular orientational disorder stabilized under increasing temperature. On the other hand, there are polymers (e.g., acetoxy silicone rubber) \cite{imamura20} and layered hybrid organic-inorganic perovskites (e.g., [C$_{10}$H$_{21}$NH$_{3}$]$_{2}$MnCl$_{4}$) \cite{lloveras21b,mason22} in which the accompanying phase transformations entail significant atomic rearrangements in the organic components. Another family of disordered materials presenting also great barocaloric promise are solid electrolytes (e.g., AgI, Li$_{3}$N and Cu$_{2}$Se) \cite{aznar17,sagotra17,sagotra18,ming20}, although in this latter case the experimentally reported $|\Delta S_{\rm rev}|$ fall slightly below the colossal threshold value of $100$~J~K$^{-1}$~kg$^{-1}$ \cite{aznar17}.

\begin{figure*}[t]
\includegraphics[width=0.80\linewidth]{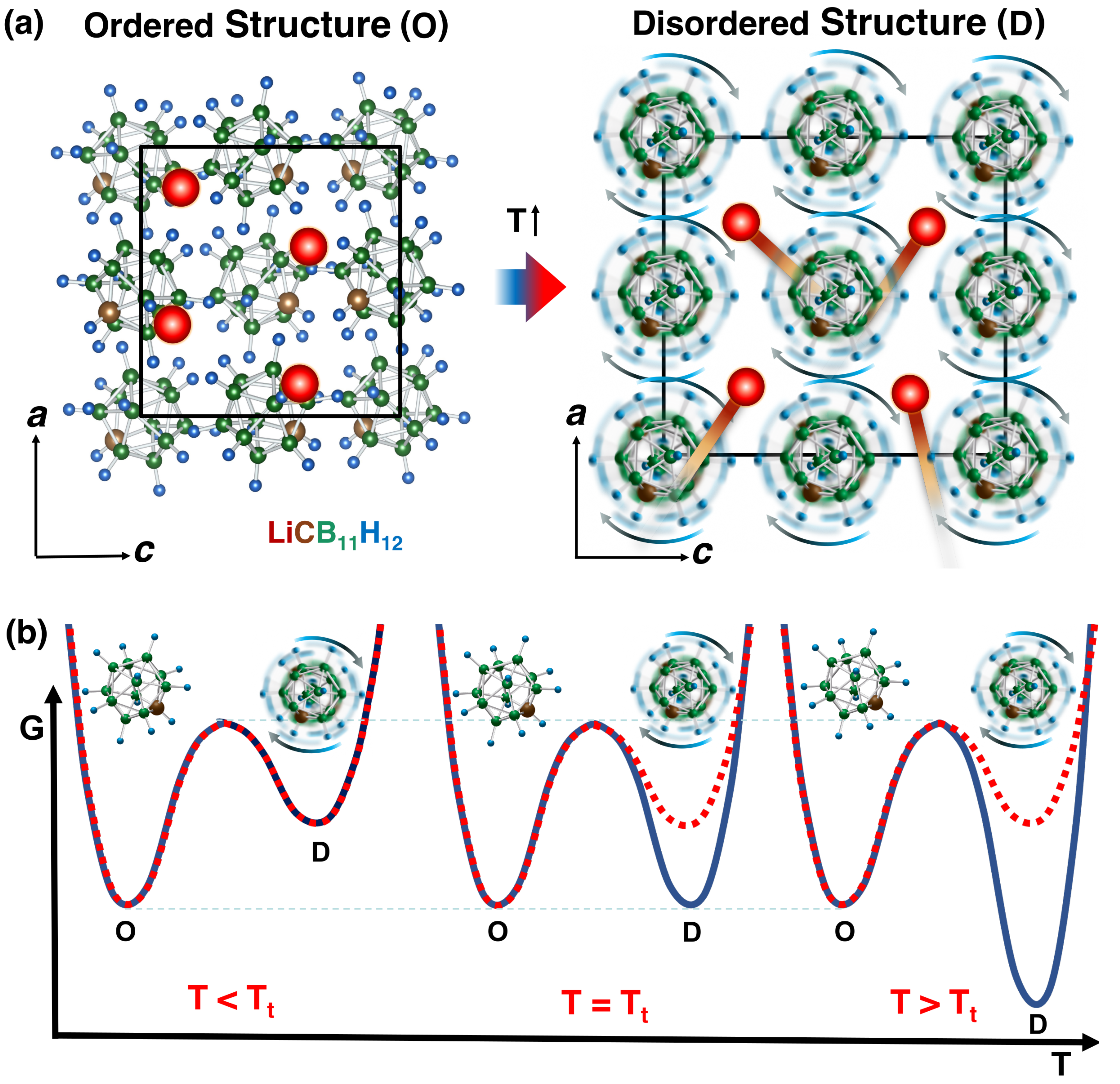}
\caption{{\bf Sketch of the order-disorder phase transition occurring in LCBH upon increasing temperature.} (a)~Ball-stick representation of the low-$T$ ordered (O) and high-$T$ disordered (D) phases. Lithium, carbon, boron and hydrogen atoms are represented with red, brown, green and blue spheres, respectively. In the high-$T$ phase, the Li$^{+}$ cations diffuse throughout the crystalline matrix while the [CB$_{11}$H$_{12}$]$^{-}$ anions reorient disorderly \cite{Tang20153637}; the volume increases significantly during the $T$-induced phase transition. (b)~Outline of the order-disorder phase transition in terms of Gibbs free energies. The red dotted lines represent internal energies and the blue solid lines Gibbs free energies; $T_{t}$ denotes the phase transition temperature.}
\label{fig1}
\end{figure*}

In spite of these recent developments, finding barocaloric materials with well-balanced and suitable features for developing thermal applications, e.g., $|\Delta T_{\rm rev}| \ge 20$~K and $|\Delta S_{\rm rev}| \ge 100$~J~K$^{-1}$~kg$^{-1}$ driven by $\Delta p \lesssim 0.1$~GPa, is proving extremely difficult. From the hundred of barocaloric materials known to date \cite{lloveras21}, to the best of our knowledge only four fulfill the conditions specified above, namely, the spin-crossover complex Fe$_3$(bntrz)$_{6}$(tcnset)$_{6}$ ($|\Delta T_{\rm rev}| = 35$~K and $|\Delta S_{\rm rev}| = 120$~J~K$^{-1}$~kg$^{-1}$ for $\Delta p = 0.26$~GPa) \cite{romanini19}, the layered hybrid perovskite [C$_{10}$H$_{21}$NH$_{3}$]$_{2}$MnCl$_{4}$ ($|\Delta T_{\rm rev}| = 27$~K and $|\Delta S_{\rm rev}| = 250$~J~K$^{-1}$~kg$^{-1}$ for $\Delta p = 0.19$~GPa) \cite{lloveras21b,mason22}, the plastic crystal 1-Br-adamantane ($|\Delta T_{\rm rev}| = 20$~K and $|\Delta S_{\rm rev}| = 120$~J~K$^{-1}$~kg$^{-1}$ for $\Delta p = 0.10$~GPa) \cite{aznar21}, and the elastomer acetoxy silicone ($|\Delta T_{\rm rev}| = 22$~K and $|\Delta S_{\rm rev}| = 182$~J~K$^{-1}$~kg$^{-1}$ for $\Delta p = 0.17$~GPa) \cite{imamura20}. Moreover, studies addressing a fundamental and quantitative understanding of the atomistic mechanisms that bring on such colossal BCE are very scarce \cite{li20,hui22,oliveira23,sau21b}, thus hindering the rational design of disordered materials with enhanced barocaloric performances.

\begin{figure*}[t]
\includegraphics[width=0.90\linewidth]{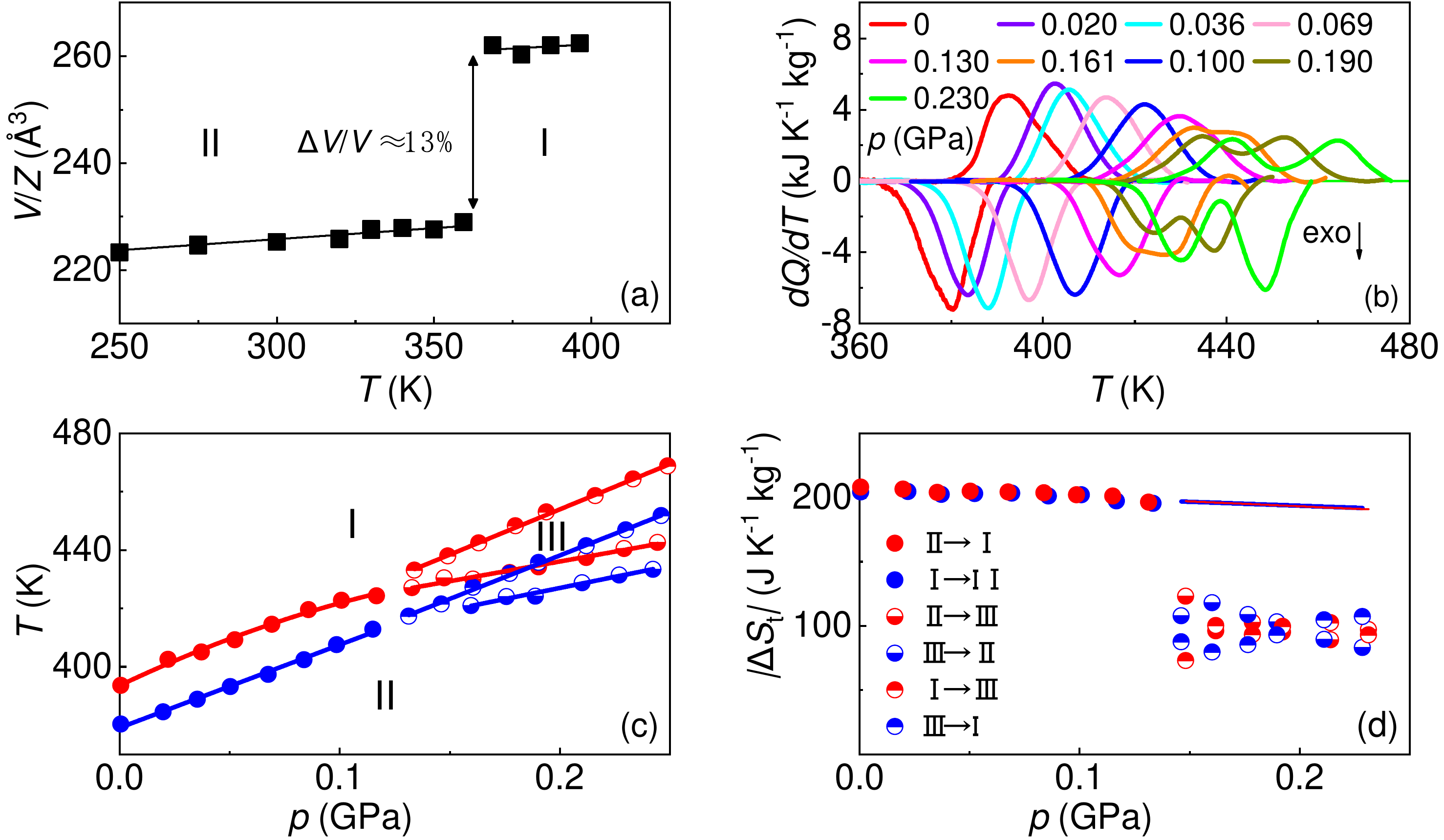}
\caption{{\bf Experimental phase diagram of bulk LCBH and corresponding phase transition entropy changes.} (a)~Volume per formula unit measured as a function of temperature at normal pressure. (b)~Isobaric heat flow data expressed as a function of applied pressure and temperature; data collected during heating (cooling) are represented in the positive (negative) y-axis. (c)~Pressure and temperature phase diagram; transition temperatures are determined from the peaks in panel (b). (d)~Phase transition entropy changes as a function of pressure and transition path. $\Delta S_\text{t}$ remains practically constant from atmospheric pressure all the way up to the triple point. At $p \simeq 0.13$~GPa, $\Delta S_{\text{II} \rightarrow \text{I}} \approx \Delta S_{\text{II} \rightarrow \text{III}}+\Delta S_{\text{III} \rightarrow \text{I}}$, while above the triple point $\Delta S_{\text{II} \rightarrow \text{III}} \approx \Delta S_{\text{III} \rightarrow \text{I}}$. Straight lines at pressures above the triple point are linear fits to $\Delta S_{\text{II} \rightarrow \text{III}}+\Delta S_{\text{III} \rightarrow \text{I}}$.}
\label{fig2}
\end{figure*}

In this work, we experimentally and theoretically demonstrate the existence of colossal and reversible BCE in the monocarba-\textit{closo}-dodecaborate LiCB$_{11}$H$_{12}$ (LCBH) near its order-disorder phase transition occurring at $T_{t} \approx 380$~K \cite{Tang20153637}. LCBH is a well-known solid electrolyte in which at temperatures above $T_{t}$ the lithium cations are highly mobile and the molecular anions [CB$_{11}$H$_{12}$]$^{-}$ reorient disorderly \cite{skripov15,Mohtadi16} (Fig.~\ref{fig1}); thus, LCBH combines phase-transition features of both plastic crystals and superionic compounds, two families of materials for which colossal and giant BCE, respectively, have been previously reported \cite{lloveras19,li19,aznar20,aznar17}. In particular, we measured colossal values of $|\Delta T_{\rm rev}| = 32$~K and $|\Delta S_{\rm rev}| = 280$~JK$^{-1}$kg$^{-1}$ for a pressure shift of $0.23$~GPa, and large and reversible barocaloric strengths of $\approx 2$~J~K$^{-1}$~kg$^{-1}$~MPa$^{-1}$ over a wide temperature interval of several tens of degrees. Likewise, for a smaller pressure shift of $0.10$~GPa assuring values of $|\Delta S_{\rm rev}| = 200$~J~K$^{-1}$~kg$^{-1}$ and $|\Delta T_{\rm rev}| = 10$~K were obtained. Atomistic molecular dynamics simulations were performed to reveal key phase transition mechanisms and quantify the role played by the vibrational, molecular orientational and ion diffusive degrees of freedom on the disclosed BCE. Very interestingly, the contribution of the lattice vibrations to $\Delta S$ was found to be the dominant at all pressures, instead of the typically assumed one resulting from molecular reorientational motion \cite{li20,hui22,oliveira23}. Our results provide new valuable insights into the physical behavior and functionality of plastic crystals and suggest that colossal BCE similar to those reported here for LCBH could also exist in other akin \textit{closo}-borate materials like NaCB$_{11}$B$_{12}$ \cite{Tang20153637,skripov15}, KCB$_{11}$B$_{12}$ \cite{udovic20}, and LiCB$_9$H$_{10}$ \cite{Kim2019,Kim20204831}.

\begin{figure*}[t]
\includegraphics[width=0.90\linewidth]{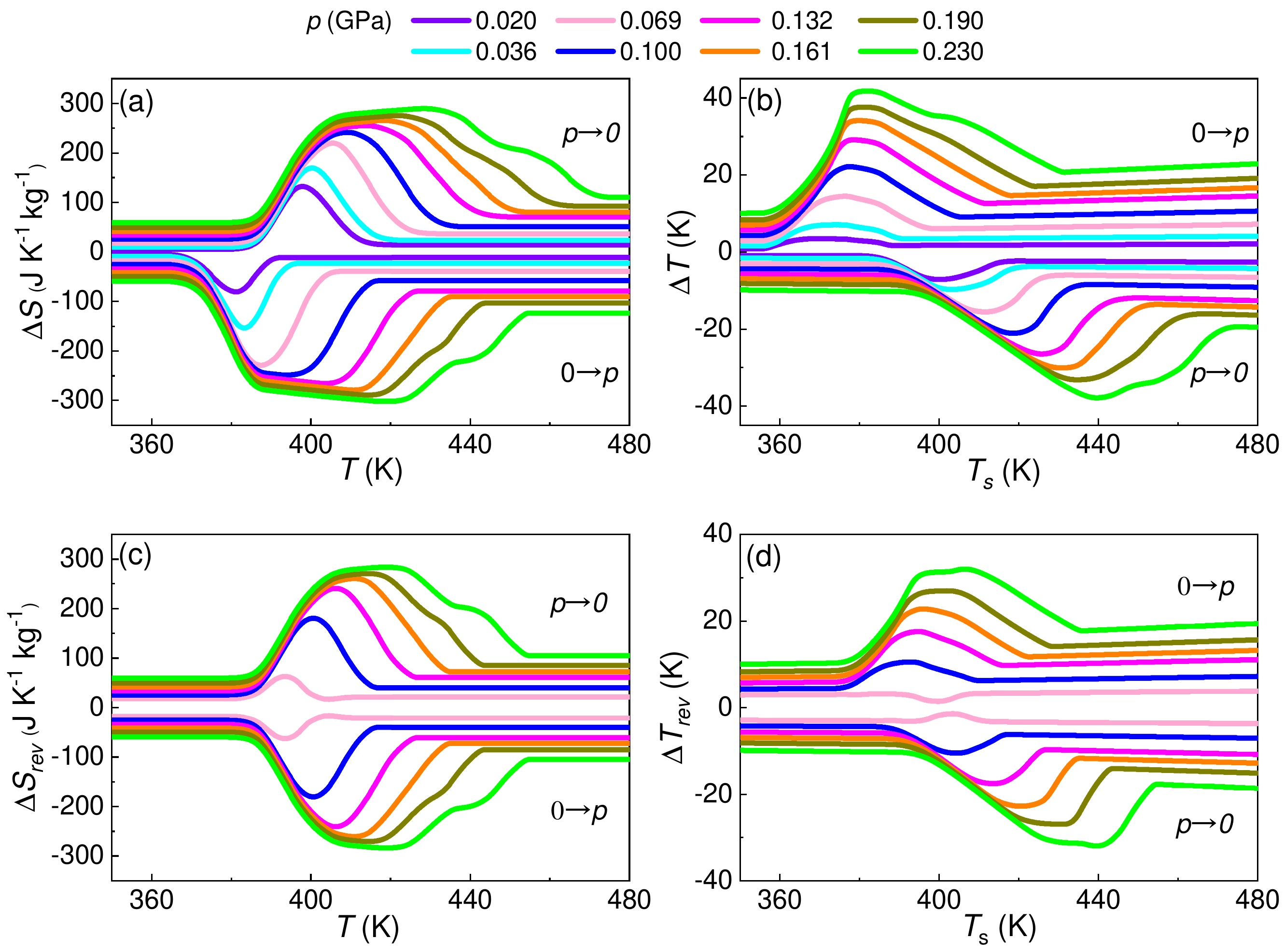}
\caption{{\bf Experimentally measured colossal barocaloric effects in bulk LCBH.} (a)--(c) Isothermal entropy change, $\Delta S$, and (b)--(d) adiabatic temperature change, $\Delta T$, obtained upon the application and removal of pressure, $p$, considering (a)--(b) irreversible and (c)--(d) reversible processes.}
\label{fig3}
\end{figure*}

\begin{figure*}[t]
\includegraphics[width=1.00\linewidth]{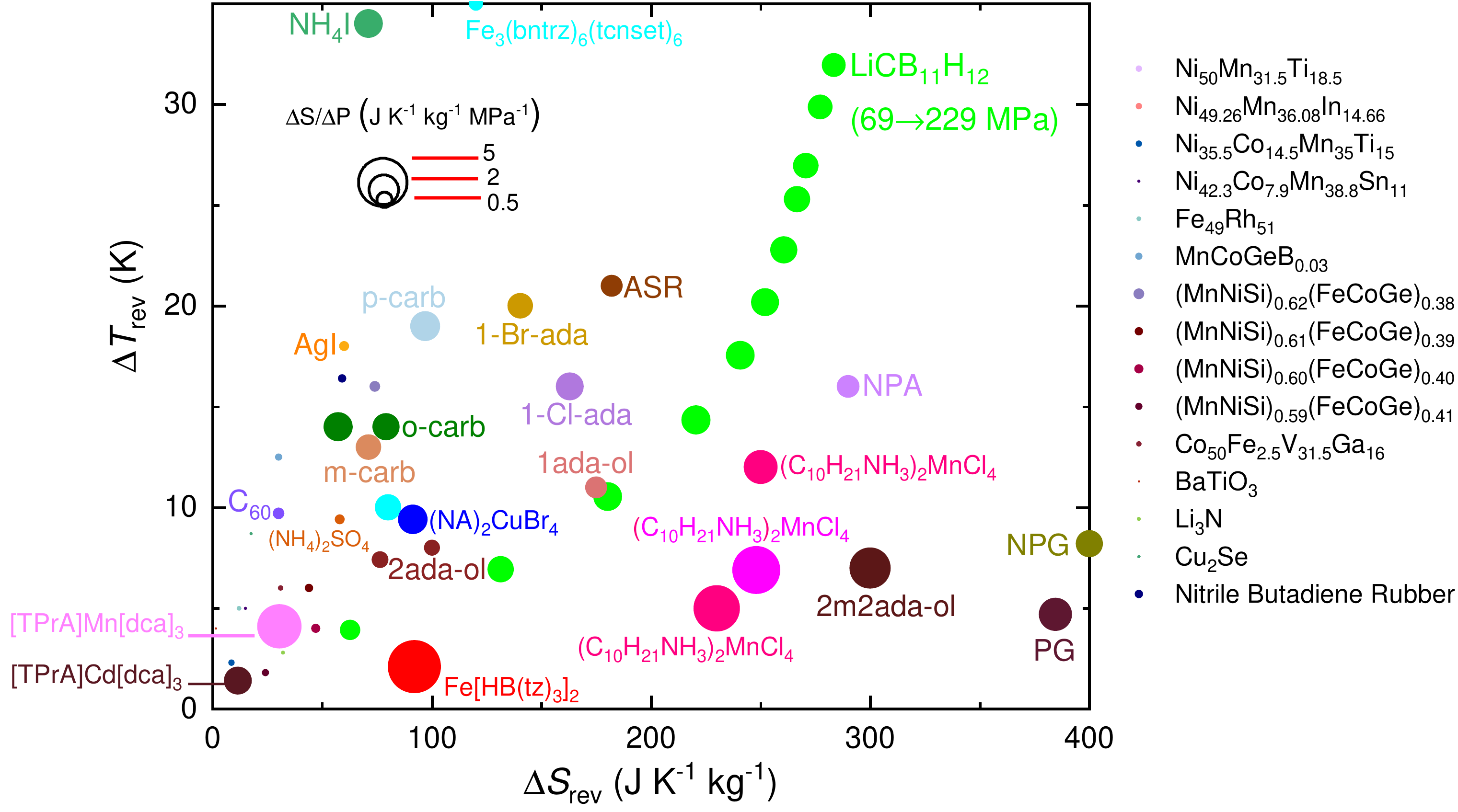}
\caption{{\bf Compendium of experimentally measured reversible BCE.} The size of the symbols represents the reversible barocaloric strength defined as the ratio of $|\Delta S_{\rm rev}|$ by the corresponding pressure change $\Delta p$. Material names are indicated near each symbol or in the right side of the panel. NPG: neopentylglycol; PG: pentaglycerine; NPA: Neopentyl alcohol; o-carb: orthocarborane; m-carb: metacarborane; p-carb: paracarborane; 1-Br-ada: 1-Bromoadamantane; 1-Cl-ada: 1-Chloroadamantane; 1ada-ol: 1-adamantanol; 2ada-ol: 2-adamantanol; 2m2ada-ol: 2-methyl-2-adamantanol; ASR: Acetoxy Silicone Rubber. Numerical details and references can be found in the Supplementary Table~S1.}
\label{fig6}
\end{figure*}

\section{Results and Discussion}
\label{sec:results}

\subsection{LiCB$_{11}$H$_{12}$ general properties}
\label{subsec:lcbh}
In a recent X-ray powder diffraction study \cite{Tang20153637}, it has been shown that at room temperature LiCB$_{11}$H$_{12}$ (LCBH) presents an ordered orthorhombic structure (space group $Pca2_{1}$) in which the Li$^{+}$ cations reside near trigonal-planar sites surrounded by molecular [CB$_{11}$H$_{12}$]$^{-}$ anions arranged in a cubic sublattice. An order-disorder phase transition occurs at $T_{t} \approx 380$~K that stabilizes a disordered phase in which the Li$^{+}$ cations are highly mobile and the molecular anions present fast reorientational motion (Fig.~\ref{fig1}a). At normal pressure, the lithium ion conductivity measured just above $T_{t}$ exceeds values of $0.1$~S cm$^{-1}$ \cite{Tang20153637} and the reorientational motion of the molecular anions can reach frequencies of $10^{11}$~s$^{-1}$ \cite{Tang20153637,Sau20212357}. Meanwhile, the $T$-induced order-disorder phase transition is accompanied by a huge volume increase of the order of $\approx 10$\% \cite{Sau20212357} that, based on the Clausius-Clapeyron (CC) equation $\Delta S_{t} = \Delta V_{t} \frac{dp}{dT}$, suggests great barocaloric potential. 

The described order-disorder phase transition can be qualitatively understood in terms of the Gibbs free energy difference between the high-$T$ disordered~(D) and low-$T$ ordered~(O) phases, $\Delta G \equiv G^{D} - G^{O}$ (Fig.~\ref{fig1}b). This free energy difference consists of an internal energy ($\Delta E$), entropy ($-T \Delta S$), and volume ($p \Delta V$) terms. The internal energy remains more or less constant during the phase transition while the volume term disfavors the stabilization of the disordered phase since $\Delta V > 0$. Thus, the LCBH order-disorder phase transition appears to be governed by the change in entropy, $\Delta S$, which in view of the ion conductivity and molecular reorientational frequency measured above $T_{t}$ should be fairly large.

\begin{figure*}[t]
\includegraphics[width=0.90\linewidth]{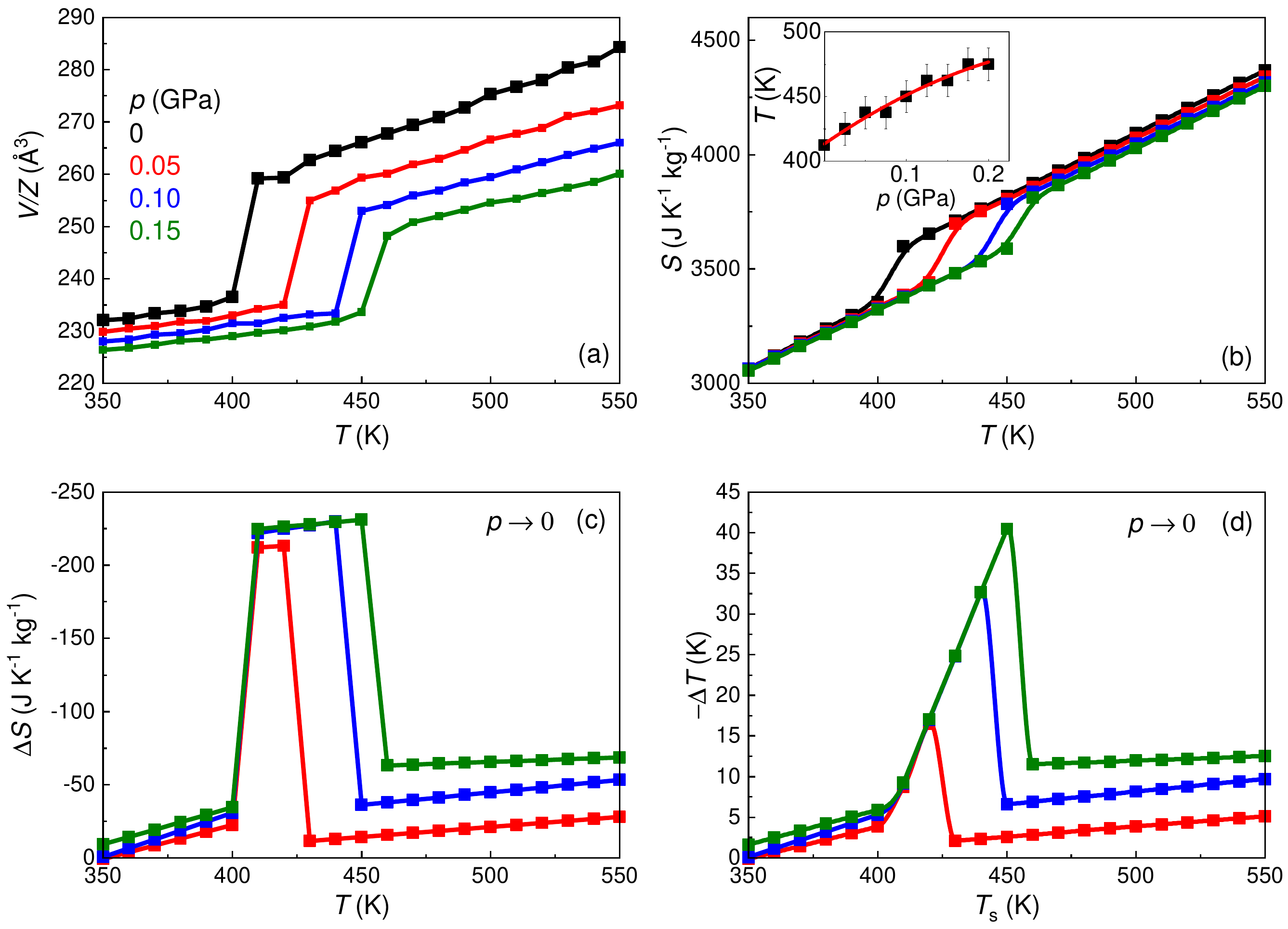}
\caption{{\bf Colossal BCE estimated for bulk LCBH with MD simulations.} (a)~Volume change per formula unit across the phase transition expressed as a function of temperature and pressure. (b)~Total entropy curves expressed as a function of pressure and temperature. \textit{Inset}: theoretically calculated $p$--$T$ phase diagram. (c)~Isothermal entropy and (d)~adiabatic temperature changes expressed as a function of temperature and pressure. Results were obtained from $NpT$-MD simulations.}
\label{fig4}
\end{figure*}

\begin{figure*}[t]
\includegraphics[width=0.90\linewidth]{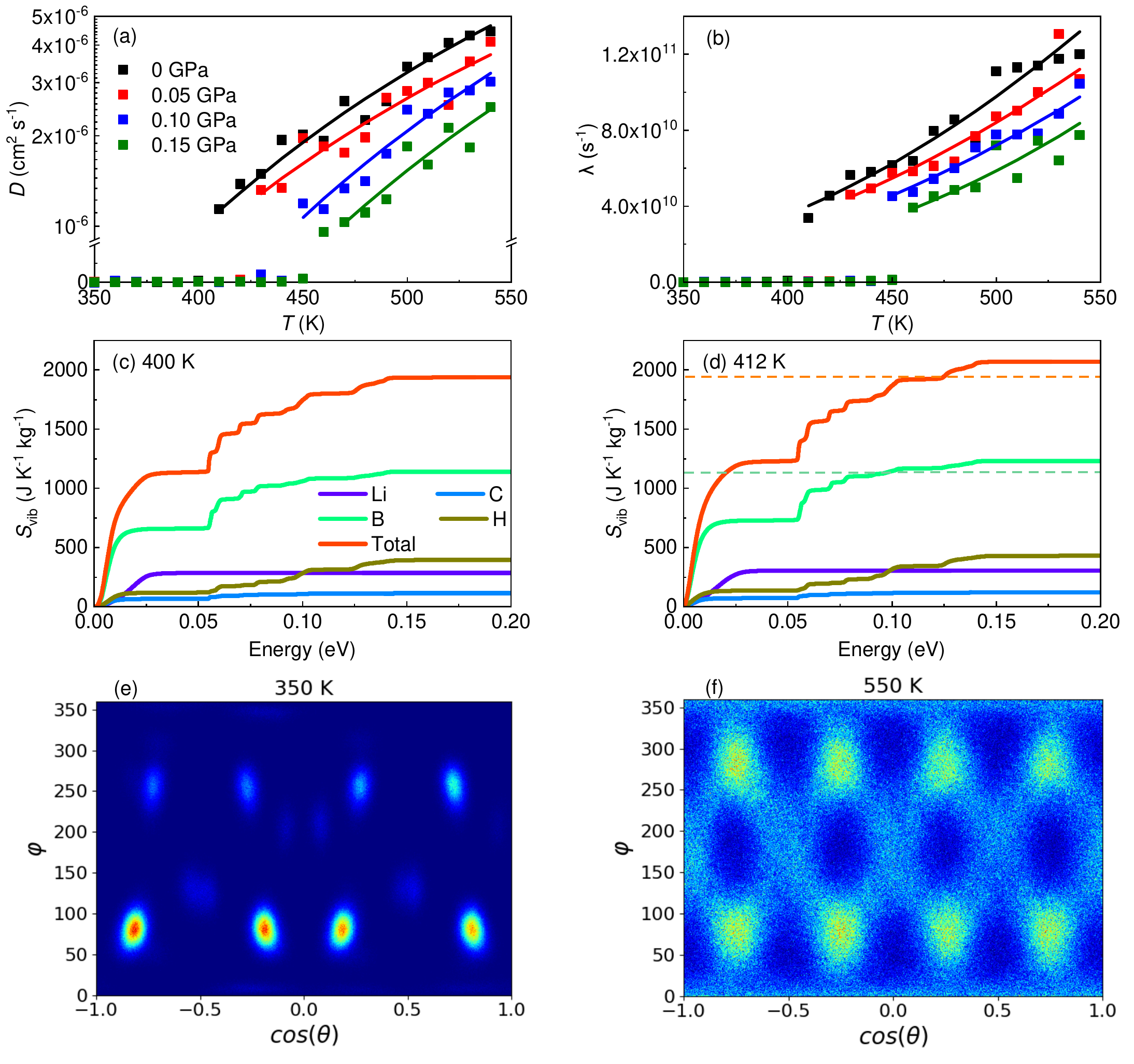}
\caption{{\bf Atomistic insights into the order-disorder phase transition in LCBH from MD simulations.} (a)~Lithium ion diffusion coefficient, $D_{\rm Li}$. (b)~Anionic reorientational frequency, $\lambda_{\rm CBH}$. Solid lines correspond to Arrhenius law fits. (c)--(d)~Cumulative function of the vibrational entropy as a function of the phonon energy and atomic species, calculated for the ordered ($T = 400$~K) and disordered ($T = 412$~K) phases at zero pressure. Dashed lines indicate analogous asymptotic values reached in the ordered phase. (e)--(f)~Angular probability density function estimated for the molecular (CB$_{11}$H$_{12}$)$^{-}$ anions calculated in the ordered ($T = 350$~K) and disordered ($T = 550$~K) phases at zero pressure, expressed as a function of the polar ($\theta$) and azimuthal ($\phi$) angles. Dark and bright areas represent low and high probability regions, respectively.}
\label{fig5}
\end{figure*}

\subsection{Experimental barocaloric results}
\label{subsec:experiments}
Conventional X-ray powder diffraction experiments performed at normal pressure and under varying temperature confirmed the expected structures of the low-$T$ and high-$T$ phases (orthorhombic and cubic symmetry, respectively). Pattern matching analysis of the obtained data yielded the temperature-dependent volume of LCBH (see Fig.~\ref{fig2}a), which shows a huge $\approx 13$\% relative volume increase at the endothermic transition corresponding to $\Delta V\approx 12\cdot10^{-5}$ m$^3$ kg$^{-1}$. 

High-pressure differential thermal  analysis (HP-DTA) was carried out in the pressure interval $0 \le p \le 0.23$~GPa (Fig.~\ref{fig2}b). At pressures below $\approx 0.13$~GPa, a single peak in the heat flow was measured corresponding to the aforementioned orthorhombic~(ordered phase, II)~$\leftrightarrow$~cubic~(disordered phase, I) first-order phase transition. At pressure above $\approx 0.13$~GPa, the HP-DTA signals exhibit two peaks thus indicating the appearance of a new phase that we label here as III (high-pressure enantiotropy). To the best of our knowledge, phase III has not been previously reported in the literature and its specific crystalline structure remains unknown since we did not resolve it. Interestingly, a broad peak was previously detected in differential scanning calorimetry experiments \cite{Tang20153637} that hints at the stabilization of phase III. 

Transition temperatures were determined from the maximum of the HP-DTA peaks (Fig.~\ref{fig2}c) which allowed to estimate an upper threshold for the triple point at $\approx$ (425 K, 0.13 GPa) given the width of the peaks obtained under the chosen scanning rate. Considering only the data measured near atmospheric pressure, the pressure dependence of the II$\rightarrow$I transition was determined to be $\frac{dT}{dp}\approx 420$ K GPa$^{-1}$, which slightly decreases under increasing pressure due to the small convexity of the coexistence line. For the II$\rightarrow$III and III$\rightarrow$I transitions, linear fits to the obtained coexistence lines yielded $\frac{dT}{dp}\approx 135$ K GPa$^{-1}$ and $\frac{dT}{dp}\approx 310$ K GPa$^{-1}$, respectively. Phase transition entropy changes were calculated via integration of the $\frac{1}{T}\frac{dQ}{dT}$ function after baseline subtraction. As it was already expected, the $\Delta S_{{\rm II} \rightarrow {\rm I}}$ values associated to the LCBH order-disorder phase transition are noticeably large, namely, $\approx208$~J~K$^{-1}$~kg$^{-1}$ (Fig.~\ref{fig2}d). By plugging the measured $\frac{dT}{dp}$ and $\Delta S_{{\rm II} \rightarrow {\rm I}}$ values at atmospheric pressure in the CC equation we obtain $\Delta V_\text{CC}\approx 9\cdot10^{-5}$ m$^3$ kg$^{-1}$, which is in reasonable agreement with the $\Delta V$ determined directly in the experiments.

Above $p\approx0.13$ GPa, due to the overlapping between the II$\leftrightarrow$III and III$\leftrightarrow$I peaks, the contribution associated to each phase transition was decided at the inflection point of the cumulative entropy change function $\int_{T_1}^{T}\frac{1}{T'}\frac{dQ}{dT'}dT'$. $\Delta S_\text{t}$ remains practically constant from atmospheric pressure all the way up to the triple point. At $p \simeq 0.13$~GPa, we obtained $\Delta S_{\text{II} \rightarrow \text{I}} \approx \Delta S_{\text{II} \rightarrow \text{III}}+\Delta S_{\text{III} \rightarrow \text{I}}$, as it is required by the condition of thermodynamic equilibrium, while above the triple point $\Delta S_{\text{II} \rightarrow \text{III}} \approx \Delta S_{\text{III} \rightarrow \text{I}}$. Splitting of the II$\rightarrow$I phase transition into II$\rightarrow$III and III$\rightarrow$I might be associated to the decoupling of the diffusive and orientational degrees of freedom right at the stabilization of the high-$T$ phase, although further investigations are necessary for a more conclusive assessment of phase III.

HP-DTA measurements along with experimental differential scanning calorimetry (Supplementary Fig.~S1), heat capacity (Supplementary Fig.~S2) and theoretical equations of state $V(T,p)$ (i.e., obtained from molecular dynamics simulations, Sec.~\ref{subsec:simulations}) were used to determine the isobaric entropy curves $S(T,p)$ (Supplementary Fig.~S3), from which the BC effects can be directly calculated (Methods). Figures~\ref{fig3}a,b show representative isothermal entropy changes, $|\Delta S|$, and adiabatic temperature changes, $|\Delta T|$, obtained upon the first application and removal of the driving pressure shift. It is worth noticing that a small $\Delta p \approx 0.03$~GPa already produced colossal values of $|\Delta S| = 100$~J~K$^{-1}$~kg$^{-1}$ and $|\Delta T| = 8$~K, and similarly $\Delta p \approx 0.08$~GPa yielded $|\Delta S| = 250$~J~K$^{-1}$~kg$^{-1}$ and $|\Delta T| = 16$~K. For the largest pressure shift considered in this study, namely, $\Delta p \approx 0.23$~GPa, the resulting $|\Delta S|$ and $|\Delta T|$ amount to $300$~J~K$^{-1}$~kg$^{-1}$ and $40$~K, respectively. 

Operation of solid-state cooling and heating devices requires cyclic application and removal of the driving fields, for which reversible caloric effects, $|\Delta S_\text{rev}|$ and $|\Delta T_\text{rev}|$, must be considered. By reversible caloric effects we mean acquitted of phase transition hysteresis effects \cite{aznar20}. The obtained results are shown in Figs.~\ref{fig3}c,d. Colossal $|\Delta S_\text{rev}|$ were already obtained for a minimum pressure shift of $\approx 0.08$~GPa. For instance, under a moderate pressure change of $\approx 0.10$~GPa LCBH renders $|\Delta S_\text{rev}| = 200$~J~K$^{-1}$~kg$^{-1}$ and $|\Delta T_\text{rev}| = 10$~K. Meanwhile, for the largest pressure shift considered in this study we measured outstanding values of $|\Delta S_\text{rev}| = 280$~J~K$^{-1}$~kg$^{-1}$ and $|\Delta T_\text{rev}| = 32$~K.

Figure~\ref{fig6} compares most of the experimental $|\Delta S_\text{rev}|$ and $|\Delta T_\text{rev}|$ reported thus far in the literature for barocaloric materials. Additionally, the size of the symbols therein account for the materials BC strength, which is defined as the ratio of $|\Delta S_\text{rev}|$ by the corresponding pressure shift ${\Delta p}$. The best performing barocaloric materials, therefore, should appear in the top right side of the panel and with the largest possible symbol area. Each material has been represented with one or two points that best illustrate their overall barocaloric performance, while for LCBH we have selected a set of barocaloric measurements. 

Although LCBH is not the best performing material in terms of a single quality, it displays an unprecedentedly well-balanced and accomplished barocaloric portfolio consisting of colossal $|\Delta S_\text{rev}|$, large $|\Delta T_\text{rev}|$ and large BC strength obtained under moderate pressure shifts of the order of $0.10$~GPa. For instance, in terms of largest $|\Delta S_\text{rev}|$ the plastic crystal neopentylglycol (NPG) emerges as the clear winner since it holds a gigantic value of $\approx 400$~J~K$^{-1}$~kg$^{-1}$ \cite{aznar20}; however, as regards $|\Delta T_\text{rev}|$ the same material becomes a poor contestant in the presence of LCBH (that is, $\approx 8$~K versus $32$~K). Likewise, the $|\Delta T_\text{rev}|$ record holder, namely, the spin-crossover complex Fe$_3$(bntrz)$_{6}$(tcnset)$_{6}$ \cite{romanini19}, presents $|\Delta S_\text{rev}|$ and BC strength values that roughly are halves of the LCBH maxima (for instance, $\approx 120$~J~K$^{-1}$~kg$^{-1}$ versus $280$~J~K$^{-1}$~kg$^{-1}$). Therefore, LCBH can be deemed as one of the most thorough and promising barocaloric materials reported to date owing to its unique parity between sizable $|\Delta S_\text{rev}|$ and $|\Delta T_\text{rev}|$ obtained under moderate pressure shifts.

\begin{figure}[t]
\includegraphics[width=1.00\linewidth]{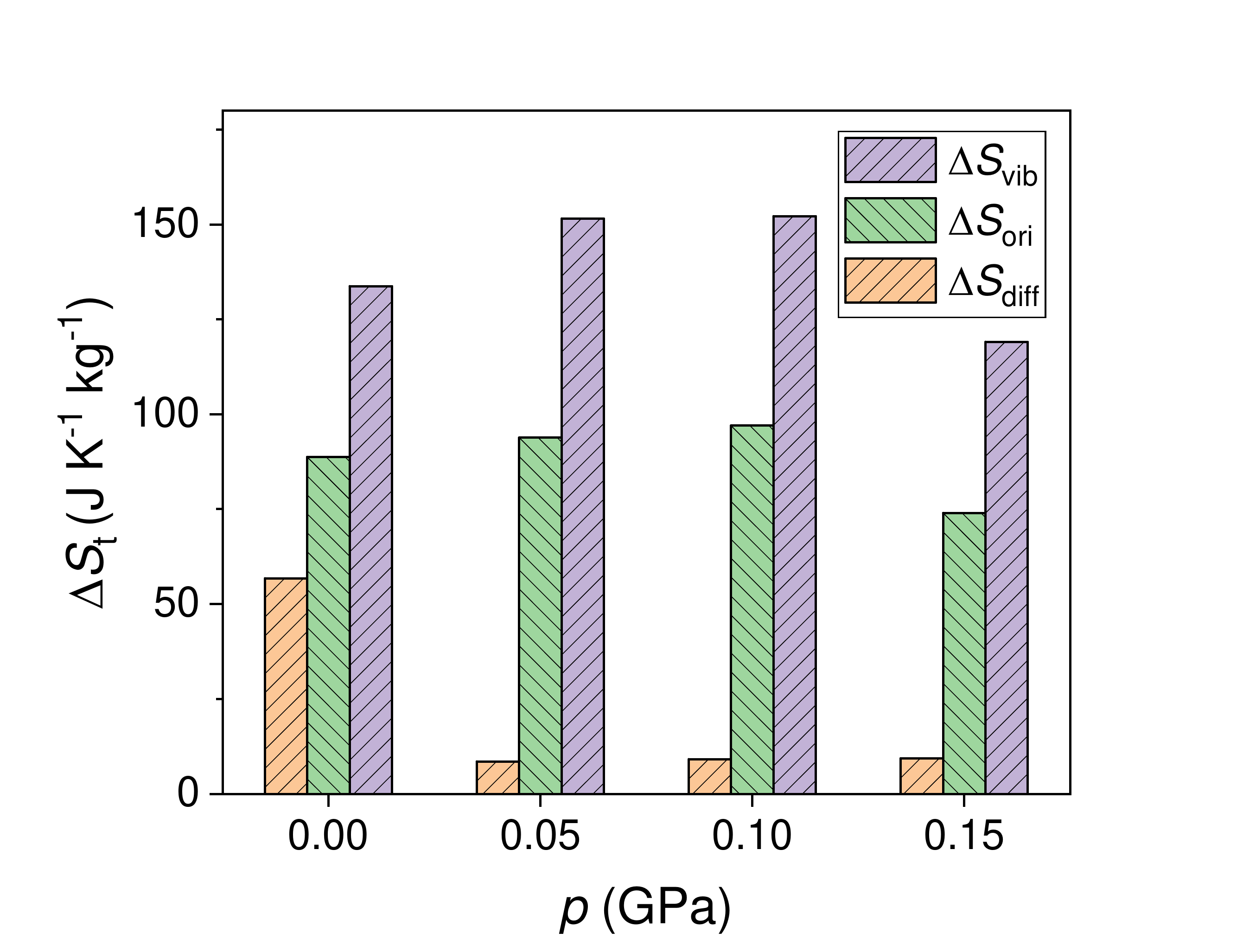}
\caption{{\bf Partial contributions to the entropy change accompanying the order-disorder phase transition in LCBH expressed as a function of pressure.} Entropy changes stem from the vibrational, $\Delta S_{\rm vib}$, molecular orientational, $\Delta S_{\rm ori}$, and cation diffusive, $\Delta S_{\rm diff}$, degrees of freedom. Results were obtained from comprehensive molecular dynamics simulations and Gibbs free energy calculations (Methods).}
\label{fig7}
\end{figure}

\subsection{Atomistic simulation of barocaloric effects}
\label{subsec:simulations}
Figures~\ref{fig4}a,b show the theoretical equation of state $V(T,p)$ and $p$--$T$ phase diagram of bulk LCBH obtained from molecular dynamics (MD) simulations (Methods). We determined the coexistence line of the high-$T$~(disordered) and low-$T$~(ordered) phases by conducting numerous MD simulations at small $p$--$T$ shifts of $0.025$~GPa and $12.5$~K. Each phase coexistence point in Fig.~\ref{fig4}b (\textit{inset}) corresponds to sharp and simultaneous changes in the volume, Li$^{+}$ diffusion coefficient ($D_{\rm Li}$), and molecular [CB$_{11}$H$_{12}$]$^{-}$ orientational frequency ($\lambda_{\rm CBH}$), as identified in the MD simulations (Figs.~\ref{fig5}a,b). At zero pressure, we estimated a huge volume increase of about $11$\% at the theoretical transition temperature $T_{t} \approx 400$~K, along with the order parameter changes $\Delta D_{\rm Li} = 1.13\cdot10^{-6}$~cm$^{2}$~s$^{-1}$ and $\Delta \lambda_{\rm CBH} = 0.33 \cdot 10^{11}$~s$^{-1}$. It was found that the pressure dependence of the transition temperature could be precisely reproduced by the second-order polynomial curve $T_{t}(p) = 412 + 438 p - 610 p^{2}$ (red line in the inset of Fig.~\ref{fig4}b), in which the temperature and pressure are expressed in units of K and GPa, respectively. The slight $\frac{dT}{dp}$ decrease under increasing compression is consistent with the $p$-induced reduction of the transition volume change since $\Delta S_{t}$ is roughly independent of pressure, in agreement with our experiments. It is worth noting that phase-transition hysteresis effects cannot be reproduced by the equilibrium MD approach employed in this study \cite{sau21b}. 

The LCBH $p$--$T$ phase diagram obtained from MD simulations (Fig.~\ref{fig4}b) is in quantitative good agreement with the experiments performed below the triple point found at $\approx 0.13$~GPa (Fig.~\ref{fig2}b), although the transition temperatures are slightly overestimated by theory. For example, at zero pressure and $p = 0.10$~GPa the MD simulations yielded $T_{t} = 410 \pm 15$ and $440 \pm 15$~K (Fig.~\ref{fig4}), respectively, to be compared with the corresponding experimental values $390 \pm 10$ and $410 \pm 10$~K (Fig.~\ref{fig2}b). The agreement between the predicted and measured volumes for the ordered and disordered phases at zero pressure is also notable, finding only small relative discrepancies of $\sim 1$\% for the low-$T$ phase (Figs.~\ref{fig2}a and \ref{fig4}a). Meanwhile, the triple point observed in the experiments was not reproduced by the MD simulations. It is worth noting, however, that under $p \neq 0$ conditions and close to $T_{t}$ we observed pre-transitional effects in our simulations consisting of few slowly diffusing Li ions in the ordered phase (Supplementary Fig.~S4). 

Figures~\ref{fig4}c,d show the theoretical barocaloric $|\Delta S|$ and $|\Delta T|$ deduced from the entropy curves $S(p,T)$ enclosed in Fig.~\ref{fig4}b, which were obtained from data generated in the MD simulations. The agreement between these theoretical results and the corresponding experimental values is remarkably good for pressures below the experimental triple point. For example, for a pressure shift of $0.10$~GPa we estimated an isothermal entropy change of $227$~J~K$^{-1}$~kg$^{-1}$ and an adiabatic temperature change of $32$~K from the MD simulations, to be compared with the corresponding experimental values $250$~J~K$^{-1}$~kg$^{-1}$ and $24$~K (Fig.~\ref{fig3}a,b). In view of such a notable agreement, we characterized with MD simulations the contributions to the phase transition entropy change stemming from the vibrational, molecular orientational and cation diffusive degrees of freedom, a highly valuable atomistic insight that in principle cannot be obtained from the experiments.  

Figures~\ref{fig5}a,b reveal synchronized surges in $D_{\rm Li}$ and $\lambda_{\rm CBH}$ at the order-disorder phase transition points. Thus, both ion diffusion and molecular anion orientational disorder (Figs.~\ref{fig5}e,f) contribute to the transition entropy change and barocaloric effects disclosed in LCBH. Nevertheless, there is a third possible source of entropy in the crystal which is related to the lattice vibrations, $S_{\rm vib}$ (Supplementary Figs.~S5-S6). Figures~\ref{fig5}c,d show examples of the cumulative $S_{\rm vib}$ function expressed as a function of the vibrational phonon energy, calculated for LCBH in the ordered and disordered phases at zero pressure and evaluated for each atomic species. Therein, it is appreciated that the largest contribution to the $S_{\rm vib}$ difference between the order and disordered phases comes from the B atoms (followed by hydrogen). This outcome can be rationalized in terms of the relative great abundance of this species in LCBH ($\approx 45$\%) and its larger mass as compared to that of H atoms ($10$ times heavier): B ions have a predominant weight on the low-frequency vibrational modes (Fig.~\ref{fig5}c-d) that most significantly contribute to $S_{\rm vib}$ near ambient temperature. 

Figure~\ref{fig7} shows the relative contributions of the vibrational, molecular orientational and ion diffusion degrees of freedom to the phase transition entropy change estimated at different pressures with MD simulations. Interestingly, in all the analyzed cases the largest contribution stems from changes in the lattice vibrations, $\Delta S_{\rm vib}$, followed by the molecular reorientations, $\Delta S_{\rm ori}$, and finally ion diffusion, $\Delta S_{\rm diff}$. For example, at zero pressure the vibrational, molecular orientational and ion diffusive degrees of freedom respectively contribute in $\approx 48$, $32$ and $20$\% to $\Delta S_{t}$. The entropy preeminence of the lattice vibrations can be rationalized in terms of (1) the huge volume expansion accompanying the order-disorder phase transition ($\sim 10$\%, Fig.~\ref{fig4}a), which further curtails the frequency of the low-energy phonon bands in the disordered phase (Supplementary Fig.~S5), and (2) the intensification and amplitude broadening of the molecular libration modes in the disordered phase (inferred from the angular probability density variations around the equilibrium positions in Figs.~\ref{fig5}e-f). These outcomes are highly valuable and insightful since thus far molecular reorientations were thought to be the primary source of entropy variation in plastic crystals undergoing order-disorder phase transitions \cite{li20,hui22,oliveira23}.   

The vibrational and orientational entropy changes remain more or less constant for pressures $\le 0.1$~GPa, whereas $\Delta S_{\rm diff}$ significantly decreases under compression. For instance, at $0.1$~GPa the diffusive degrees of freedom contribute to $\Delta S_{t}$ in less than $4$\%. These outcomes can be understood in terms of the small fraction of diffusive ions in LCBH (i.e., one Li atom per formula unit) and the marked decline in $D_{\rm Li}$ induced by pressure (Fig.~\ref{fig5}a). The appearance of pre-transitional effects in our MD simulations, specially under $p \neq 0$ conditions (Supplementary Fig.~S4), also contributes to the noticeable $\Delta S_{\rm diff}$ drop caused by compression. Nonetheless, it is worth noting that despite the relative minuteness of $\Delta S_{\rm diff}$, cation disorder was found to play a critical role on triggering molecular orientational disorder, which by contrast contributes very significantly to $\Delta S_{t}$. In particular, we conducted constrained MD runs in which we fixed the positions of the lithium ions so that they could not diffuse. It was found then that molecular orientational disorder only emerged at temperatures well above $550$~K (Supplementary Fig.~S7). Therefore, it can be concluded that cation disorder crucially assists on the realization of colossal BCE through the order-disorder phase transition, a characteristic trait that differentiates LCBH from other molecular plastic crystals bearing also great barocaloric promise.

\section{Conclusions}
\label{sec:conclusions}
Colossal barocaloric effects (BCE) driven by pressure shifts of the order of $0.10$~GPa were experimentally and theoretically disclosed in bulk LiCB$_{11}$H$_{12}$ (LCBH), a compound that at high temperatures presents disorder features characteristic of both plastic crystals and superionic materials, namely, molecular reorientational motion and ion diffusion. Reversible peaks of $|\Delta S_{\rm rev}| = 280$~J~K$^{-1}$~kg$^{-1}$ and $|\Delta T_{\rm rev}| = 32$~K were experimentally measured around $400$~K for a pressure shift of $0.23$~GPa, yielding huge and reversible barocaloric strengths of $\approx 2$~J~K$^{-1}$~kg$^{-1}$~MPa$^{-1}$ over tens of degrees intervals. Likewise, for a smaller pressure shift of $0.10$~GPa we obtained very promising values of $|\Delta S_{\rm rev}| = 200$~J~K$^{-1}$~kg$^{-1}$ and $|\Delta T_{\rm rev}| = 10$~K. These results place LCBH among the best-known barocaloric materials in terms of huge and reversible isothermal entropy and adiabatic temperature changes, two quantities that rarely are found simultaneously in a same material. 

Atomistic molecular dynamics simulations yielded theoretical $|\Delta S|$ and $|\Delta T|$ in very good agreement with the experimental values, and allowed to quantify the importance of vibrational, molecular orientational, and ion diffusive degrees of freedom on the disclosed colossal BCE. It was found that the contribution to the phase transition entropy change stemming from the lattice vibrations was the largest, followed by that of molecular reorientations and both being much superior than the entropy associated to lithium diffusion alone. Nevertheless, cationic disorder was found to have a critical influence on the stabilization of orientational disorder thus, in spite of its small contribution to $\Delta S_{t}$, lithium diffusion appears to be essential for the emergence of colossal BCE in bulk LCBH. These results are of high significance since reveal the preeminence of the vibrational degrees of freedom in the phase transition entropy change of a plastic crystal, and demonstrate atomistic BCE mechanisms other than molecular reorientational disorder (i.e., lattice vibrations and ion diffusion).  

LCBH belongs to the family of \textit{closo}-borate materials, a promising class of solid electrolytes for all-solid-state batteries. Examples of akin compounds that have been already synthesized in the laboratory and tested for electrochemical energy storage applications are NaCB$_{11}$H$_{12}$ \cite{Tang20153637,skripov15}, KCB$_{11}$H$_{12}$ \cite{udovic20}, and LiCB$_9$H$_{10}$ \cite{Kim2019,Kim20204831}. Colossal BCE could also exist in these materials and in other similar compounds harboring both ion diffusion and molecular orientational disorder at or near room temperature. Thus, the present combined experimental-theoretical study opens new horizons in solid-state cooling and heating and advances knowledge in the realization of colossal BCE in plastic crystals.

\section*{Methods}
\label{methods}
\subsection*{Experimental techniques}
\textit{Materials synthesis.}~LiCB$_{11}$H$_{12}$ was obtained by drying the hydrated compound LiCB$_{11}$H$_{12}$$\cdot$xH$_2$O (Katchem, Ltd.) under vacuum ($<5\times10^{-4}$ Pa) at 160~$^\circ$C for 12~h.
\\
\textit{X-ray powder diffraction.}~High-resolution X-ray powder diffraction measurements were performed using the Debye–Scherrer geometry and transmission mode with a horizontally mounted cylindrical position-sensitive INEL detector (CPS-120). Monochromatic Cu-K$\alpha_1$ radiation was selected by means of a curved germanium monochromator. Temperature-dependent measurements were performed using a liquid nitrogen 700 series Oxford Cryostream Cooler. Powder samples were introduced into 0.5 mm diameter Lindemann capillaries. Volume was obtained by pattern matching procedure.
\\
\textit{Quasi-direct barocaloric measurements.}~A Q100 thermal analyzer (TA Instruments) was used to perform differential scanning calorimetry experiments at atmospheric pressure with $\sim 10$~mg of sample hermetically encapsulated in Aluminum pans (Supplementary Fig.~S1). The standard mode (at 3, 5 and 10~K~min$^{-1}$) was used to determine the transition properties whereas the modulated mode (isothermal conditions, modulation amplitude 1~$^\circ$C, modulation period $120$~s) was used to measure the heat capacity in each phase (Supplementary Fig.~S2).

Pressure-dependent calorimetry was performed with a custom-built high-pressure differential thermal analyzer (from Irimo, Bellota Herramientas S.A.) that uses Bridgman thermocouples as thermal sensors. The nominal operational pressure range is from atmospheric to $0.3$~GPa and the temperature range is from room temperature up to $473$~K. Heating ramps were performed at $3$~K~min$^{-1}$ using a resistive heater whereas cooling were carried out at $\sim -2$~K~min$^{-1}$ by an air stream. A few hundreds of mg of LiCB$_{11}$H$_{12}$ were mixed with an inert perfluorinated fluid (Galden Bioblock Scientist) to remove air and sealed within tin capsules. The pressure-transmitting fluid was Therm240 (Lauda).

Isobaric entropy functions $S(T,p)$ were determined with respect to a reference temperature $T_0$ below the transition using the method explained in Ref.\cite{Li2020reversible} (Supplementary Fig.~S3). The procedure is based on the following thermodynamic equation:
\begin{equation}
S(T,p)=S(T_{0},p)+\int_{T_{0}}^T\frac{1}{T}\left(C_{p}+\frac{dQ}{dT}\right)dT~,
\end{equation}
where $\frac{dQ}{dT}$ is the heat flow in temperature due to the first-order phase transition measured by pressure-dependent calorimetry.

In each phase, $C_p$ is the corresponding heat capacity and was considered independent of pressure as indicated by the approximately linear behavior of volume with temperature obtained in the two phases from MD simulations (Fig.~\ref{fig4}) along with the thermodynamic equation:
\begin{equation}
\left(\frac{\partial C_p}{\partial p}\right)_T=-T\left(\frac{\partial^2 V}{\partial T^2}\right)_p~.
\end{equation}
In the transition region $C_p$ was calculated as an average weighted according to the fraction of each phase. To take into account the dependence of the transition region with pressure, the overall $C_p$ function at atmospheric pressure obtained in each phase and across the transition was extrapolated to higher temperatures according to the experimental value of $\frac{dT}{dp}\Delta p$, where $\Delta p$ is the pressure change applied in each particular case. Experimental measurement of $C_p$ at atmospheric pressure and the calculated curves at different pressures are shown in Supplementary Fig.~S2.

The pressure dependence of $S(T,p)$ was evaluated using the thermodynamic equation:
\begin{equation}
S(T,p)=S(T,p_{0})-\int_{p_{0}}^{p}\left(\frac{\partial V}{\partial T}\right)_{T,p'} dp'~,
\end{equation}
where $p_{0}$ was selected equal to $p_{\text{atm}} = 1$~bar. Here, we make use of the approximation $\left(\frac{\partial V}{\partial T}\right)_{T,p} \simeq \left(\frac{\partial V}{\partial T}\right)_{T,p_{0}}$, which is reasonable based on the $\left(\frac{\partial V}{\partial T}\right)_{T, p}$ data obtained from the MD simulations (Fig.~\ref{fig4}).

Once the entropy function $S(T,p)$ was determined for both heating and cooling runs independently (Supplementary Fig.~S3), BC effects obtained upon first application or removal of the field were calculated as:
\begin{eqnarray}
\Delta S(T,p_0\rightarrow p_1)=S(T,p_1)-S(T,p_0)~{\rm and}\\
\Delta T(T_s,p_0\rightarrow p_1)=T(S,p_1)-T_s(T,p_0)~,
\end{eqnarray}
where $T_s$ is the starting temperature of the heating/cooling process. Here, it must be considered that for materials with $\frac{dT}{dp} > 0$ BC effects on compression ($p_0 = p_\text{atm}$, $p_1 > p_\text{atm}$) and decompression ($p_0 > p_\text{atm}$, $p_1 = p_\text{atm}$) are calculated from $S(T,p)$ functions obtained on cooling and heating, respectively \cite{aznar20}. In turn, BC effects obtained reversibly on cyclic compression-decompression processes were calculated from the $S(T,p)$ curves obtained on heating at atmospheric pressure and cooling at high pressure.
\\

\subsection*{Simulation techniques}
\textit{Molecular dynamics simulations.}~Force-field based molecular dynamics (MD) simulations were performed using a previously reported interatomic potential for LCBH \cite{Sau20212357}. This force field is a combination of Coulomb-Buckingham (CB), harmonic bond, and angle-type potentials, namely:
\begin{subequations}
\begin{align}\label{eq_1}
    U (r, \theta) = U_{\rm{CB}} (r) + U_{\rm{bond}} (r) + U_{\rm{angle}} (\theta)~, \\
    U_{\rm{CB}} (r) = \frac{q_iq_j}{4\pi\epsilon_0 r} + A_{ij}\exp(-r/\rho)-\frac{C_{ij}}{r^6}~,\\
    U_{\rm{bond}} (r) = \frac{1}{2}k_r(r-r_0)^2~{\rm and}\\
    U_{\rm{angle}} (\theta) = \frac{1}{2}k_\theta(\theta-\theta_0)^2~,
    \end{align}
\end{subequations}
where $q_i$ denotes the charge of the ion labeled $i$, $\epsilon_0$ the vacuum permittivity,  $A_{ij}$ and $\rho$ the short-range repulsive energy and length scales for the pairs of atoms $ij$, and $C_{ij}$ the corresponding dispersion interaction coefficient. $r_0$ and $\theta_0$ are an equilibrium bond distance and angle, respectively, and $k_r$ and $k_\theta$ the spring constants of the harmonic bond and angle potentials. The numerical value of these potential parameters can be found in the Supplementary Table~S2.

We performed $NpT$-MD simulations in the temperature range $325 \le T \le 525$~K at intervals of $12.5$~K, and pressure range $0 \le p \le 0.15$~GPa at intervals of $0.025$~GPa. The temperature and pressure in the system were controlled with thermostating and barostating techniques, in which some dynamic variables are coupled with the particle velocities and simulation box dimensions. The simulation supercell comprised a total of $6400$ atoms. A time step of $0.5$~fs was employed for integration of the atomic forces along with the velocity Verlet algorithm. A typical $NpT$-MD run lasted for about $2$~ns and the atomic trajectories were stored at intervals of $500$~fs. Detailed analyses and statistical time averages were performed over the last $1$~ns of such simulations. To guarantee proper convergence of the estimated thermodynamic properties, in few instances longer simulation times of $10$~ns were carried out. Periodic boundary conditions were applied along the three Cartesian directions and the Ewald summation technique was used for evaluation of the long-range Coulomb interactions with a short-range cut-off distance of $13$~\AA. All the $NpT$-MD simulations were carried out with the LAMMPS software package \cite{lammps}.
\\

\textit{Density functional theory and ab initio molecular dynamics simulations.}~First-principles calculations based on density functional theory (DFT) were performed to analyze the energy, structural and vibrational properties of bulk LCBH. The DFT calculations were carried out with the VASP code \cite{kresse1993ab} by following the generalized gradient approximation to the exchange-correlation energy due to Perdew \textit{et al.} (PBE) \cite{Perdew19963865}. The projector augmented-wave method was used to represent the ionic cores \cite{Blochl199417953}, and the electronic states $1s$–$2s$ Li, $2s$–$2p$ C, $2s$–$2p$ B and $1s$ H were considered as valence. Wave functions were represented in a plane-wave basis truncated at $650$~eV. By using these parameters and
dense k-point grids for Brillouin zone integration, the resulting energies were converged to within $1$~meV per formula unit. In the geometry relaxations, a tolerance of $0.005$~eV~\AA$^{-1}$ was imposed in the atomic forces. 

\textit{Ab initio} molecular dynamics (AIMD) simulations based on DFT were carried out to assess the reliability of the interatomic potential model employed in the MD simulations on the description of the vibrational degrees of freedom of bulk LCBH (Supplementary Fig.~S6). The AIMD simulations were performed in the canonical ensemble $(N, V, T)$ considering constant number of particles, volume and temperature. The constrained volumes were equal to the equilibrium volumes determined at zero temperature, an approximation that has been shown to be reasonable at moderate temperatures \cite{Sagotra2019}. The temperature in the AIMD simulations was kept fluctuating around a set-point value by using Nose-Hoover thermostats. A large simulation box containing 800 atoms was employed in all the simulations, and periodic boundary conditions were applied along the three Cartesian directions. Newton’s equations of motion were integrated by using the customary Verlet’s algorithm and a time-step length of $\delta t = 10^{-3}$~ps. $\Gamma$-point sampling for integration within the first Brillouin zone was employed in all the AIMD simulations. The AIMD simulations comprised long simulation times of $\approx 200$~ps and temperatures in the range $200 \le T \le 500$~K.
\\

\textit{Estimation of key quantities with MD simulations.}~The mean square displacement of the lithium ions was estimated with the formula \cite{cazorla19b}:
\begin{eqnarray}
{\rm MSD_{\rm Li}}(\tau) & = & \frac{1}{N_{\rm ion} \left( N_{\rm step} - n_{\tau} \right)} \times \\ \nonumber
                &   & \sum_{i=1}^{N_{\rm ion}} \sum_{j=1}^{N_{\rm step} - n_{\tau}} | {\bf r}_{i} (t_{j} + \tau) - {\bf r}_{i} (t_{j}) |^{2}~, 
\label{eq1}
\end{eqnarray}
where ${\bf r}_{i}(t_{j})$ is the position of the migrating ion $i$ at time $t_{j}$ ($= j \cdot \delta t$), $\tau$ represents a lag time, $n_{\tau} = \tau / \delta t$, $N_{\rm ion}$ is the total number of mobile ions, and $N_{\rm step}$ the total number of time steps. The maximum $n_{\tau}$ was chosen equal to $N_{\rm step}/2$, hence we could accumulate enough statistics to reduce significantly the fluctuations in ${\rm MSD_{\rm Li}}(\tau)$ at large $\tau$'s. The diffusion coefficient of lithium ions was calculated with the Einstein's relation:
\begin{equation}
D_{\rm Li} =  \lim_{\tau \to \infty} \frac{{\rm MSD_{Li}}(\tau)}{6\tau}~,  
\label{eq2}     
\end{equation}
by performing linear fits to the averaged ${\rm MSD_{Li}}$ values calculated at long $\tau$.

The angular autocorrelation function of the molecular [CB$_{11}$H$_{12}$]$^{-}$ anions was estimated using the expression \cite{sau21b}:
\begin{equation}
        \phi_{\rm CBH} (\tau) = \langle \hat{{\bf r}} (t) \cdot \hat{{\bf r}} (t + \tau) \rangle~,
\label{eq3}
\end{equation}
where $\hat{{\bf r}}$ is a unitary vector connecting the center of mass of each closoborane unit with one of its edges and $\langle \cdots \rangle$ denotes statistical average in the $(N,p,T)$ ensemble considering all the molecular anions. This autocorrelation function typically decays as $\propto \exp{[-\lambda_{\rm CBH} \cdot \tau]}$, where the parameter $\lambda_{\rm CBH}$ represents a characteristic reorientational frequency. For significant anion reorientational motion, that is, large $\lambda_{\rm CBH}$, the $\phi_{\rm CBH}$ function decreases rapidly to zero with time. 

The temperature dependence of the lithium diffusion coefficient was assumed to follow an Arrhenius law at any pressure of the form:
\begin{equation}
D_{\rm Li} (T) = D_{0} \cdot e^{-(\frac{E_a}{k_{B}T})}~,
\label{arrhenius}
\end{equation}
where $D_{0}$ and $E_{a}$ are parameters that depend on $p$ and $k_{B}$ represents the Boltzmann constant. The reorientational frequency of closoborane units, $\lambda_{\rm CBH}$, was assumed to follow a similar dependence on temperature. 

The entropy of each phase was calculated as a function of temperature and pressure, $S(p,T)$, by fully considering the vibrational, molecular orientational and ion diffusive degrees of freedom:
\begin{equation}
    S(p,T) = S_{\rm vib}(p,T) + S_{\rm ori}(p,T) + S_{\rm diff}(p,T)~.
\end{equation}
In the low-$T$ phase, $S_{\rm ori}$ and $S_{\rm diff}$ are null while in the high-$T$ phase are finite and positive. 

The vibrational density of states (VDOS), $g(\omega)$, was calculated via the Fourier transform of the velocity-velocity autocorrelation function obtained directly from the $NpT$-MD simulations, namely:
\begin{equation}
        g(\omega) = \frac{1}{N_{ion}} \sum_{i}^{N_{ion}} \int_{0}^{\infty} 
        \langle {\bf v}_{i}(\tau)\cdot{\bf v}_{i}(0)\rangle e^{i\omega \tau} d\tau~,
\label{eq5}     
\end{equation}
where ${\bf v}_{i}(t)$ represents the velocity of the atom labeled $i$ at time $t$, and $\langle \cdots \rangle$ denotes statistical average in the $(N,p,T)$ ensemble. The vibrational entropy was subsequently estimated with the formula \cite{qha}:
\begin{eqnarray}
S_{\rm vib} (p,T) & = & -\int_0^{\infty} k_{B}\ln{\left[2\sinh{\left(\frac{\hbar\omega}{2k_BT}\right)}\right]} \hat{g}(\omega) d\omega + \nonumber \\ 
& & \int_0^{\infty} \frac{\hbar\omega}{2T} \tanh^{-1}{\left(\frac{\hbar\omega}{2k_BT}\right)} \hat{g}(\omega) d\omega~,
\end{eqnarray}
where $\hat{g}(\omega)$ is the normalized vibrational density of states ($\int_{0}^{\infty} \hat{g}(\omega) d\omega = 3 N_{ion}$) and the dependence on pressure (and also temperature) is implicitly contained in $\hat{g}(\omega)$.

The orientational entropy of the molecular anions, $S_{\rm ori}$, was directly calculated from the angular probability density, $\rho(\theta, \phi)$, like \cite{takagi20}:
\begin{equation}
    S_{\rm ori}(p,T) = -k_{B}\int_{0}^{\pi} \int_{0}^{2\pi} \rho(\theta, \phi) \ln{\rho(\theta, \phi)}~ d\theta d\phi~,
\end{equation}
where $\rho(\theta, \phi)$ was obtained from the $NpT$-MD simulation runs in the form of average histograms (Fig.~\ref{fig5}). 

The ion diffusive entropy difference was estimated at the phase transition points via equalization of the Gibbs free energies of the low-$T$~(O) and high-$T$~(D) phases, namely, $G^{D}(p,T_{t}) = G^{O}(p,T_{t})$, thus leading to the expression: 
\begin{equation}
\Delta S_{\rm diff}(p,T_{t}) = \frac{\langle \Delta E \rangle}{T_{t}} + p\frac{\langle \Delta V \rangle}{T_{t}} -\Delta S_{\rm vib} - \Delta S_{\rm ori}~,
\end{equation}
where $\Delta X \equiv X^{D} - X^{O}$ and $E$ represents the internal energy of the system. For any pressure, $\Delta S_{\rm diff}$ was assumed to be constant at temperatures $T_{t} \le T$.
\\

\bibliography{MD.bib}

\providecommand{\noopsort}[1]{}\providecommand{\singleletter}[1]{#1}%
\begin{thebibliography}{40}%
\makeatletter
\providecommand \@ifxundefined [1]{%
 \@ifx{#1\undefined}
}%
\providecommand \@ifnum [1]{%
 \ifnum #1\expandafter \@firstoftwo
 \else \expandafter \@secondoftwo
 \fi
}%
\providecommand \@ifx [1]{%
 \ifx #1\expandafter \@firstoftwo
 \else \expandafter \@secondoftwo
 \fi
}%
\providecommand \natexlab [1]{#1}%
\providecommand \enquote  [1]{``#1''}%
\providecommand \bibnamefont  [1]{#1}%
\providecommand \bibfnamefont [1]{#1}%
\providecommand \citenamefont [1]{#1}%
\providecommand \href@noop [0]{\@secondoftwo}%
\providecommand \href [0]{\begingroup \@sanitize@url \@href}%
\providecommand \@href[1]{\@@startlink{#1}\@@href}%
\providecommand \@@href[1]{\endgroup#1\@@endlink}%
\providecommand \@sanitize@url [0]{\catcode `\\12\catcode `\$12\catcode
  `\&12\catcode `\#12\catcode `\^12\catcode `\_12\catcode `\%12\relax}%
\providecommand \@@startlink[1]{}%
\providecommand \@@endlink[0]{}%
\providecommand \url  [0]{\begingroup\@sanitize@url \@url }%
\providecommand \@url [1]{\endgroup\@href {#1}{\urlprefix }}%
\providecommand \urlprefix  [0]{URL }%
\providecommand \Eprint [0]{\href }%
\providecommand \doibase [0]{http://dx.doi.org/}%
\providecommand \selectlanguage [0]{\@gobble}%
\providecommand \bibinfo  [0]{\@secondoftwo}%
\providecommand \bibfield  [0]{\@secondoftwo}%
\providecommand \translation [1]{[#1]}%
\providecommand \BibitemOpen [0]{}%
\providecommand \bibitemStop [0]{}%
\providecommand \bibitemNoStop [0]{.\EOS\space}%
\providecommand \EOS [0]{\spacefactor3000\relax}%
\providecommand \BibitemShut  [1]{\csname bibitem#1\endcsname}%
\let\auto@bib@innerbib\@empty
\bibitem [{\citenamefont {Moya}\ and\ \citenamefont {Mathur}(2020)}]{Moya2020}%
  \BibitemOpen
  \bibfield  {author} {\bibinfo {author} {\bibfnamefont {X.}~\bibnamefont
  {Moya}}\ and\ \bibinfo {author} {\bibfnamefont {N.~D.}\ \bibnamefont
  {Mathur}},\ }\href {\doibase 10.1126/science.abb0973} {\bibfield  {journal}
  {\bibinfo  {journal} {Science}\ }\textbf {\bibinfo {volume} {370}},\ \bibinfo
  {pages} {797} (\bibinfo {year} {2020})}\BibitemShut {NoStop}%
\bibitem [{\citenamefont {Mañosa}\ \emph {et~al.}(2013)\citenamefont
  {Mañosa}, \citenamefont {Planes},\ and\ \citenamefont {Acet}}]{manosa13}%
  \BibitemOpen
  \bibfield  {author} {\bibinfo {author} {\bibfnamefont {L.}~\bibnamefont
  {Mañosa}}, \bibinfo {author} {\bibfnamefont {A.}~\bibnamefont {Planes}}, \
  and\ \bibinfo {author} {\bibfnamefont {M.}~\bibnamefont {Acet}},\ }\href
  {\doibase 10.1039/C3TA01289A} {\bibfield  {journal} {\bibinfo  {journal}
  {Journal of Materials Chemistry A}\ }\textbf {\bibinfo {volume} {1}},\
  \bibinfo {pages} {4925} (\bibinfo {year} {2013})}\BibitemShut {NoStop}%
\bibitem [{\citenamefont {Hou}\ \emph {et~al.}(2022)\citenamefont {Hou},
  \citenamefont {Qian},\ and\ \citenamefont {Takeuchi}}]{Hou2022}%
  \BibitemOpen
  \bibfield  {author} {\bibinfo {author} {\bibfnamefont {H.}~\bibnamefont
  {Hou}}, \bibinfo {author} {\bibfnamefont {S.}~\bibnamefont {Qian}}, \ and\
  \bibinfo {author} {\bibfnamefont {I.}~\bibnamefont {Takeuchi}},\ }\href
  {\doibase 10.1038/s41578-022-00428-x} {\bibfield  {journal} {\bibinfo
  {journal} {Nature Reviews Materials}\ }\textbf {\bibinfo {volume} {7}},\
  \bibinfo {pages} {633} (\bibinfo {year} {2022})}\BibitemShut {NoStop}%
\bibitem [{\citenamefont {Mañosa}\ and\ \citenamefont
  {Planes}(2017)}]{manosa17}%
  \BibitemOpen
  \bibfield  {author} {\bibinfo {author} {\bibfnamefont {L.}~\bibnamefont
  {Mañosa}}\ and\ \bibinfo {author} {\bibfnamefont {A.}~\bibnamefont
  {Planes}},\ }\href {\doibase https://doi.org/10.1002/adma.201603607}
  {\bibfield  {journal} {\bibinfo  {journal} {Advanced Materials}\ }\textbf
  {\bibinfo {volume} {29}},\ \bibinfo {pages} {1603607} (\bibinfo {year}
  {2017})}\BibitemShut {NoStop}%
\bibitem [{\citenamefont {Cazorla}(2019)}]{cazorla19}%
  \BibitemOpen
  \bibfield  {author} {\bibinfo {author} {\bibfnamefont {C.}~\bibnamefont
  {Cazorla}},\ }\href {\doibase 10.1063/1.5113620} {\bibfield  {journal}
  {\bibinfo  {journal} {Applied Physics Reviews}\ }\textbf {\bibinfo {volume}
  {6}},\ \bibinfo {pages} {041316} (\bibinfo {year} {2019})}\BibitemShut
  {NoStop}%
\bibitem [{\citenamefont {Lloveras}\ and\ \citenamefont
  {Tamarit}(2021)}]{lloveras21}%
  \BibitemOpen
  \bibfield  {author} {\bibinfo {author} {\bibfnamefont {P.}~\bibnamefont
  {Lloveras}}\ and\ \bibinfo {author} {\bibfnamefont {J.-L.}\ \bibnamefont
  {Tamarit}},\ }\href {\doibase 10.1557/s43581-020-00002-4} {\bibfield
  {journal} {\bibinfo  {journal} {MRS Energy {\&} Sustainability}\ }\textbf
  {\bibinfo {volume} {8}},\ \bibinfo {pages} {3} (\bibinfo {year}
  {2021})}\BibitemShut {NoStop}%
\bibitem [{\citenamefont {Lloveras}\ \emph {et~al.}(2019)\citenamefont
  {Lloveras}, \citenamefont {Aznar}, \citenamefont {Barrio}, \citenamefont
  {Negrier}, \citenamefont {Popescu}, \citenamefont {Planes}, \citenamefont
  {Ma{\~{n}}osa}, \citenamefont {Stern-Taulats}, \citenamefont {Avramenko},
  \citenamefont {Mathur}, \citenamefont {Moya},\ and\ \citenamefont
  {Tamarit}}]{lloveras19}%
  \BibitemOpen
  \bibfield  {author} {\bibinfo {author} {\bibfnamefont {P.}~\bibnamefont
  {Lloveras}}, \bibinfo {author} {\bibfnamefont {A.}~\bibnamefont {Aznar}},
  \bibinfo {author} {\bibfnamefont {M.}~\bibnamefont {Barrio}}, \bibinfo
  {author} {\bibfnamefont {P.}~\bibnamefont {Negrier}}, \bibinfo {author}
  {\bibfnamefont {C.}~\bibnamefont {Popescu}}, \bibinfo {author} {\bibfnamefont
  {A.}~\bibnamefont {Planes}}, \bibinfo {author} {\bibfnamefont
  {L.}~\bibnamefont {Ma{\~{n}}osa}}, \bibinfo {author} {\bibfnamefont
  {E.}~\bibnamefont {Stern-Taulats}}, \bibinfo {author} {\bibfnamefont
  {A.}~\bibnamefont {Avramenko}}, \bibinfo {author} {\bibfnamefont {N.~D.}\
  \bibnamefont {Mathur}}, \bibinfo {author} {\bibfnamefont {X.}~\bibnamefont
  {Moya}}, \ and\ \bibinfo {author} {\bibfnamefont {J.-L.}\ \bibnamefont
  {Tamarit}},\ }\href {\doibase 10.1038/s41467-019-09730-9} {\bibfield
  {journal} {\bibinfo  {journal} {Nature Communications}\ }\textbf {\bibinfo
  {volume} {10}},\ \bibinfo {pages} {1803} (\bibinfo {year}
  {2019})}\BibitemShut {NoStop}%
\bibitem [{\citenamefont {Li}\ \emph {et~al.}(2019)\citenamefont {Li},
  \citenamefont {Kawakita}, \citenamefont {Ohira-Kawamura}, \citenamefont
  {Sugahara}, \citenamefont {Wang}, \citenamefont {Wang}, \citenamefont {Chen},
  \citenamefont {Kawaguchi}, \citenamefont {Kawaguchi}, \citenamefont {Ohara},
  \citenamefont {Li}, \citenamefont {Yu}, \citenamefont {Mole}, \citenamefont
  {Hattori}, \citenamefont {Kikuchi}, \citenamefont {Yano}, \citenamefont
  {Zhang}, \citenamefont {Zhang}, \citenamefont {Ren}, \citenamefont {Lin},
  \citenamefont {Sakata}, \citenamefont {Nakajima},\ and\ \citenamefont
  {Zhang}}]{li19}%
  \BibitemOpen
  \bibfield  {author} {\bibinfo {author} {\bibfnamefont {B.}~\bibnamefont
  {Li}}, \bibinfo {author} {\bibfnamefont {Y.}~\bibnamefont {Kawakita}},
  \bibinfo {author} {\bibfnamefont {S.}~\bibnamefont {Ohira-Kawamura}},
  \bibinfo {author} {\bibfnamefont {T.}~\bibnamefont {Sugahara}}, \bibinfo
  {author} {\bibfnamefont {H.}~\bibnamefont {Wang}}, \bibinfo {author}
  {\bibfnamefont {J.}~\bibnamefont {Wang}}, \bibinfo {author} {\bibfnamefont
  {Y.}~\bibnamefont {Chen}}, \bibinfo {author} {\bibfnamefont {S.~I.}\
  \bibnamefont {Kawaguchi}}, \bibinfo {author} {\bibfnamefont {S.}~\bibnamefont
  {Kawaguchi}}, \bibinfo {author} {\bibfnamefont {K.}~\bibnamefont {Ohara}},
  \bibinfo {author} {\bibfnamefont {K.}~\bibnamefont {Li}}, \bibinfo {author}
  {\bibfnamefont {D.}~\bibnamefont {Yu}}, \bibinfo {author} {\bibfnamefont
  {R.}~\bibnamefont {Mole}}, \bibinfo {author} {\bibfnamefont {T.}~\bibnamefont
  {Hattori}}, \bibinfo {author} {\bibfnamefont {T.}~\bibnamefont {Kikuchi}},
  \bibinfo {author} {\bibfnamefont {S.-i.}\ \bibnamefont {Yano}}, \bibinfo
  {author} {\bibfnamefont {Z.}~\bibnamefont {Zhang}}, \bibinfo {author}
  {\bibfnamefont {Z.}~\bibnamefont {Zhang}}, \bibinfo {author} {\bibfnamefont
  {W.}~\bibnamefont {Ren}}, \bibinfo {author} {\bibfnamefont {S.}~\bibnamefont
  {Lin}}, \bibinfo {author} {\bibfnamefont {O.}~\bibnamefont {Sakata}},
  \bibinfo {author} {\bibfnamefont {K.}~\bibnamefont {Nakajima}}, \ and\
  \bibinfo {author} {\bibfnamefont {Z.}~\bibnamefont {Zhang}},\ }\href
  {\doibase 10.1038/s41586-019-1042-5} {\bibfield  {journal} {\bibinfo
  {journal} {Nature}\ }\textbf {\bibinfo {volume} {567}},\ \bibinfo {pages}
  {506} (\bibinfo {year} {2019})}\BibitemShut {NoStop}%
\bibitem [{\citenamefont {Aznar}\ \emph {et~al.}(2020)\citenamefont {Aznar},
  \citenamefont {Lloveras}, \citenamefont {Barrio}, \citenamefont {Negrier},
  \citenamefont {Planes}, \citenamefont {Mañosa}, \citenamefont {Mathur},
  \citenamefont {Moya},\ and\ \citenamefont {Tamarit}}]{aznar20}%
  \BibitemOpen
  \bibfield  {author} {\bibinfo {author} {\bibfnamefont {A.}~\bibnamefont
  {Aznar}}, \bibinfo {author} {\bibfnamefont {P.}~\bibnamefont {Lloveras}},
  \bibinfo {author} {\bibfnamefont {M.}~\bibnamefont {Barrio}}, \bibinfo
  {author} {\bibfnamefont {P.}~\bibnamefont {Negrier}}, \bibinfo {author}
  {\bibfnamefont {A.}~\bibnamefont {Planes}}, \bibinfo {author} {\bibfnamefont
  {L.}~\bibnamefont {Mañosa}}, \bibinfo {author} {\bibfnamefont {N.~D.}\
  \bibnamefont {Mathur}}, \bibinfo {author} {\bibfnamefont {X.}~\bibnamefont
  {Moya}}, \ and\ \bibinfo {author} {\bibfnamefont {J.-L.}\ \bibnamefont
  {Tamarit}},\ }\href {\doibase 10.1039/C9TA10947A} {\bibfield  {journal}
  {\bibinfo  {journal} {Journal of Materials Chemistry A}\ }\textbf {\bibinfo
  {volume} {8}},\ \bibinfo {pages} {639} (\bibinfo {year} {2020})}\BibitemShut
  {NoStop}%
\bibitem [{\citenamefont {Aznar}\ \emph {et~al.}(2021)\citenamefont {Aznar},
  \citenamefont {Negrier}, \citenamefont {Planes}, \citenamefont {Mañosa},
  \citenamefont {Stern-Taulats}, \citenamefont {Moya}, \citenamefont {Barrio},
  \citenamefont {Tamarit},\ and\ \citenamefont {Lloveras}}]{aznar21}%
  \BibitemOpen
  \bibfield  {author} {\bibinfo {author} {\bibfnamefont {A.}~\bibnamefont
  {Aznar}}, \bibinfo {author} {\bibfnamefont {P.}~\bibnamefont {Negrier}},
  \bibinfo {author} {\bibfnamefont {A.}~\bibnamefont {Planes}}, \bibinfo
  {author} {\bibfnamefont {L.}~\bibnamefont {Mañosa}}, \bibinfo {author}
  {\bibfnamefont {E.}~\bibnamefont {Stern-Taulats}}, \bibinfo {author}
  {\bibfnamefont {X.}~\bibnamefont {Moya}}, \bibinfo {author} {\bibfnamefont
  {M.}~\bibnamefont {Barrio}}, \bibinfo {author} {\bibfnamefont {J.-L.}\
  \bibnamefont {Tamarit}}, \ and\ \bibinfo {author} {\bibfnamefont
  {P.}~\bibnamefont {Lloveras}},\ }\href {\doibase
  https://doi.org/10.1016/j.apmt.2021.101023} {\bibfield  {journal} {\bibinfo
  {journal} {Applied Materials Today}\ }\textbf {\bibinfo {volume} {23}},\
  \bibinfo {pages} {101023} (\bibinfo {year} {2021})}\BibitemShut {NoStop}%
\bibitem [{\citenamefont {Zhang}\ \emph {et~al.}(2022)\citenamefont {Zhang},
  \citenamefont {Song}, \citenamefont {Qi}, \citenamefont {Zhang},
  \citenamefont {Zhang}, \citenamefont {Yu}, \citenamefont {Li}, \citenamefont
  {Zhang},\ and\ \citenamefont {Li}}]{li22}%
  \BibitemOpen
  \bibfield  {author} {\bibinfo {author} {\bibfnamefont {K.}~\bibnamefont
  {Zhang}}, \bibinfo {author} {\bibfnamefont {R.}~\bibnamefont {Song}},
  \bibinfo {author} {\bibfnamefont {J.}~\bibnamefont {Qi}}, \bibinfo {author}
  {\bibfnamefont {Z.}~\bibnamefont {Zhang}}, \bibinfo {author} {\bibfnamefont
  {Z.}~\bibnamefont {Zhang}}, \bibinfo {author} {\bibfnamefont
  {C.}~\bibnamefont {Yu}}, \bibinfo {author} {\bibfnamefont {K.}~\bibnamefont
  {Li}}, \bibinfo {author} {\bibfnamefont {Z.}~\bibnamefont {Zhang}}, \ and\
  \bibinfo {author} {\bibfnamefont {B.}~\bibnamefont {Li}},\ }\href {\doibase
  https://doi.org/10.1002/adfm.202112622} {\bibfield  {journal} {\bibinfo
  {journal} {Advanced Functional Materials}\ }\textbf {\bibinfo {volume}
  {32}},\ \bibinfo {pages} {2112622} (\bibinfo {year} {2022})}\BibitemShut
  {NoStop}%
\bibitem [{\citenamefont {Imamura}\ \emph {et~al.}(2020)\citenamefont
  {Imamura}, \citenamefont {Usuda}, \citenamefont {Paix{\~a}o}, \citenamefont
  {Bom}, \citenamefont {Gomes},\ and\ \citenamefont {Carvalho}}]{imamura20}%
  \BibitemOpen
  \bibfield  {author} {\bibinfo {author} {\bibfnamefont {W.}~\bibnamefont
  {Imamura}}, \bibinfo {author} {\bibfnamefont {{\'E}.~O.}\ \bibnamefont
  {Usuda}}, \bibinfo {author} {\bibfnamefont {L.~S.}\ \bibnamefont
  {Paix{\~a}o}}, \bibinfo {author} {\bibfnamefont {N.~M.}\ \bibnamefont {Bom}},
  \bibinfo {author} {\bibfnamefont {A.~M.}\ \bibnamefont {Gomes}}, \ and\
  \bibinfo {author} {\bibfnamefont {A.~M.~G.}\ \bibnamefont {Carvalho}},\
  }\href {\doibase 10.1007/s10118-020-2423-9} {\bibfield  {journal} {\bibinfo
  {journal} {Chinese Journal of Polymer Science}\ }\textbf {\bibinfo {volume}
  {38}},\ \bibinfo {pages} {999} (\bibinfo {year} {2020})}\BibitemShut
  {NoStop}%
\bibitem [{\citenamefont {Li}\ \emph {et~al.}(2021)\citenamefont {Li},
  \citenamefont {Barrio}, \citenamefont {Dunstan}, \citenamefont {Dixey},
  \citenamefont {Lou}, \citenamefont {Tamarit}, \citenamefont {Phillips},\ and\
  \citenamefont {Lloveras}}]{lloveras21b}%
  \BibitemOpen
  \bibfield  {author} {\bibinfo {author} {\bibfnamefont {J.}~\bibnamefont
  {Li}}, \bibinfo {author} {\bibfnamefont {M.}~\bibnamefont {Barrio}}, \bibinfo
  {author} {\bibfnamefont {D.~J.}\ \bibnamefont {Dunstan}}, \bibinfo {author}
  {\bibfnamefont {R.}~\bibnamefont {Dixey}}, \bibinfo {author} {\bibfnamefont
  {X.}~\bibnamefont {Lou}}, \bibinfo {author} {\bibfnamefont {J.-L.}\
  \bibnamefont {Tamarit}}, \bibinfo {author} {\bibfnamefont {A.~E.}\
  \bibnamefont {Phillips}}, \ and\ \bibinfo {author} {\bibfnamefont
  {P.}~\bibnamefont {Lloveras}},\ }\href {\doibase
  https://doi.org/10.1002/adfm.202105154} {\bibfield  {journal} {\bibinfo
  {journal} {Advanced Functional Materials}\ }\textbf {\bibinfo {volume}
  {31}},\ \bibinfo {pages} {2105154} (\bibinfo {year} {2021})}\BibitemShut
  {NoStop}%
\bibitem [{\citenamefont {Seo}\ \emph {et~al.}(2022)\citenamefont {Seo},
  \citenamefont {McGillicuddy}, \citenamefont {Slavney}, \citenamefont {Zhang},
  \citenamefont {Ukani}, \citenamefont {Yakovenko}, \citenamefont {Zheng},\
  and\ \citenamefont {Mason}}]{mason22}%
  \BibitemOpen
  \bibfield  {author} {\bibinfo {author} {\bibfnamefont {J.}~\bibnamefont
  {Seo}}, \bibinfo {author} {\bibfnamefont {R.~D.}\ \bibnamefont
  {McGillicuddy}}, \bibinfo {author} {\bibfnamefont {A.~H.}\ \bibnamefont
  {Slavney}}, \bibinfo {author} {\bibfnamefont {S.}~\bibnamefont {Zhang}},
  \bibinfo {author} {\bibfnamefont {R.}~\bibnamefont {Ukani}}, \bibinfo
  {author} {\bibfnamefont {A.~A.}\ \bibnamefont {Yakovenko}}, \bibinfo {author}
  {\bibfnamefont {S.-L.}\ \bibnamefont {Zheng}}, \ and\ \bibinfo {author}
  {\bibfnamefont {J.~A.}\ \bibnamefont {Mason}},\ }\href {\doibase
  10.1038/s41467-022-29800-9} {\bibfield  {journal} {\bibinfo  {journal}
  {Nature Communications}\ }\textbf {\bibinfo {volume} {13}},\ \bibinfo {pages}
  {2536} (\bibinfo {year} {2022})}\BibitemShut {NoStop}%
\bibitem [{\citenamefont {Salvatori}\ \emph {et~al.}(2022)\citenamefont
  {Salvatori}, \citenamefont {Negrier}, \citenamefont {Aznar}, \citenamefont
  {Barrio}, \citenamefont {Tamarit},\ and\ \citenamefont
  {Lloveras}}]{Salvatori2022}%
  \BibitemOpen
  \bibfield  {author} {\bibinfo {author} {\bibfnamefont {A.}~\bibnamefont
  {Salvatori}}, \bibinfo {author} {\bibfnamefont {P.}~\bibnamefont {Negrier}},
  \bibinfo {author} {\bibfnamefont {A.}~\bibnamefont {Aznar}}, \bibinfo
  {author} {\bibfnamefont {M.}~\bibnamefont {Barrio}}, \bibinfo {author}
  {\bibfnamefont {J.~L.}\ \bibnamefont {Tamarit}}, \ and\ \bibinfo {author}
  {\bibfnamefont {P.}~\bibnamefont {Lloveras}},\ }\href@noop {} {\bibfield
  {journal} {\bibinfo  {journal} {APL Materials}\ }\textbf {\bibinfo {volume}
  {10}},\ \bibinfo {pages} {111117} (\bibinfo {year} {2022})}\BibitemShut
  {NoStop}%
\bibitem [{\citenamefont {Aznar}\ \emph {et~al.}(2017)\citenamefont {Aznar},
  \citenamefont {Lloveras}, \citenamefont {Romanini}, \citenamefont {Barrio},
  \citenamefont {Tamarit}, \citenamefont {Cazorla}, \citenamefont {Errandonea},
  \citenamefont {Mathur}, \citenamefont {Planes}, \citenamefont {Moya},\ and\
  \citenamefont {Ma{\~{n}}osa}}]{aznar17}%
  \BibitemOpen
  \bibfield  {author} {\bibinfo {author} {\bibfnamefont {A.}~\bibnamefont
  {Aznar}}, \bibinfo {author} {\bibfnamefont {P.}~\bibnamefont {Lloveras}},
  \bibinfo {author} {\bibfnamefont {M.}~\bibnamefont {Romanini}}, \bibinfo
  {author} {\bibfnamefont {M.}~\bibnamefont {Barrio}}, \bibinfo {author}
  {\bibfnamefont {J.-L.}\ \bibnamefont {Tamarit}}, \bibinfo {author}
  {\bibfnamefont {C.}~\bibnamefont {Cazorla}}, \bibinfo {author} {\bibfnamefont
  {D.}~\bibnamefont {Errandonea}}, \bibinfo {author} {\bibfnamefont {N.~D.}\
  \bibnamefont {Mathur}}, \bibinfo {author} {\bibfnamefont {A.}~\bibnamefont
  {Planes}}, \bibinfo {author} {\bibfnamefont {X.}~\bibnamefont {Moya}}, \ and\
  \bibinfo {author} {\bibfnamefont {L.}~\bibnamefont {Ma{\~{n}}osa}},\ }\href
  {\doibase 10.1038/s41467-017-01898-2} {\bibfield  {journal} {\bibinfo
  {journal} {Nature Communications}\ }\textbf {\bibinfo {volume} {8}},\
  \bibinfo {pages} {1851} (\bibinfo {year} {2017})}\BibitemShut {NoStop}%
\bibitem [{\citenamefont {Sagotra}\ \emph {et~al.}(2017)\citenamefont
  {Sagotra}, \citenamefont {Errandonea},\ and\ \citenamefont
  {Cazorla}}]{sagotra17}%
  \BibitemOpen
  \bibfield  {author} {\bibinfo {author} {\bibfnamefont {A.~K.}\ \bibnamefont
  {Sagotra}}, \bibinfo {author} {\bibfnamefont {D.}~\bibnamefont {Errandonea}},
  \ and\ \bibinfo {author} {\bibfnamefont {C.}~\bibnamefont {Cazorla}},\ }\href
  {\doibase 10.1038/s41467-017-01081-7} {\bibfield  {journal} {\bibinfo
  {journal} {Nature Communications}\ }\textbf {\bibinfo {volume} {8}},\
  \bibinfo {pages} {963} (\bibinfo {year} {2017})}\BibitemShut {NoStop}%
\bibitem [{\citenamefont {Sagotra}\ \emph {et~al.}(2018)\citenamefont
  {Sagotra}, \citenamefont {Chu},\ and\ \citenamefont {Cazorla}}]{sagotra18}%
  \BibitemOpen
  \bibfield  {author} {\bibinfo {author} {\bibfnamefont {A.~K.}\ \bibnamefont
  {Sagotra}}, \bibinfo {author} {\bibfnamefont {D.}~\bibnamefont {Chu}}, \ and\
  \bibinfo {author} {\bibfnamefont {C.}~\bibnamefont {Cazorla}},\ }\href
  {\doibase 10.1038/s41467-018-05835-9} {\bibfield  {journal} {\bibinfo
  {journal} {Nature Communications}\ }\textbf {\bibinfo {volume} {9}},\
  \bibinfo {pages} {3337} (\bibinfo {year} {2018})}\BibitemShut {NoStop}%
\bibitem [{\citenamefont {Min}\ \emph {et~al.}(2020)\citenamefont {Min},
  \citenamefont {Sagotra},\ and\ \citenamefont {Cazorla}}]{ming20}%
  \BibitemOpen
  \bibfield  {author} {\bibinfo {author} {\bibfnamefont {J.}~\bibnamefont
  {Min}}, \bibinfo {author} {\bibfnamefont {A.~K.}\ \bibnamefont {Sagotra}}, \
  and\ \bibinfo {author} {\bibfnamefont {C.}~\bibnamefont {Cazorla}},\ }\href
  {\doibase 10.1103/PhysRevMaterials.4.015403} {\bibfield  {journal} {\bibinfo
  {journal} {Phys. Rev. Materials}\ }\textbf {\bibinfo {volume} {4}},\ \bibinfo
  {pages} {015403} (\bibinfo {year} {2020})}\BibitemShut {NoStop}%
\bibitem [{\citenamefont {Tang}\ \emph {et~al.}(2015)\citenamefont {Tang},
  \citenamefont {Unemoto}, \citenamefont {Zhou}, \citenamefont {Stavila},
  \citenamefont {Matsuo}, \citenamefont {Wu}, \citenamefont {Orimo},\ and\
  \citenamefont {Udovic}}]{Tang20153637}%
  \BibitemOpen
  \bibfield  {author} {\bibinfo {author} {\bibfnamefont {W.}~\bibnamefont
  {Tang}}, \bibinfo {author} {\bibfnamefont {A.}~\bibnamefont {Unemoto}},
  \bibinfo {author} {\bibfnamefont {W.}~\bibnamefont {Zhou}}, \bibinfo {author}
  {\bibfnamefont {V.}~\bibnamefont {Stavila}}, \bibinfo {author} {\bibfnamefont
  {M.}~\bibnamefont {Matsuo}}, \bibinfo {author} {\bibfnamefont
  {H.}~\bibnamefont {Wu}}, \bibinfo {author} {\bibfnamefont {S.-I.}\
  \bibnamefont {Orimo}}, \ and\ \bibinfo {author} {\bibfnamefont
  {T.}~\bibnamefont {Udovic}},\ }\href {\doibase 10.1039/c5ee02941d} {\bibfield
   {journal} {\bibinfo  {journal} {Energy and Environmental Science}\ }\textbf
  {\bibinfo {volume} {8}},\ \bibinfo {pages} {3637} (\bibinfo {year}
  {2015})}\BibitemShut {NoStop}%
\bibitem [{\citenamefont {Romanini}\ \emph {et~al.}(2021)\citenamefont
  {Romanini}, \citenamefont {Wang}, \citenamefont {Gürpinar}, \citenamefont
  {Ornelas}, \citenamefont {Lloveras}, \citenamefont {Zhang}, \citenamefont
  {Zheng}, \citenamefont {Barrio}, \citenamefont {Aznar}, \citenamefont
  {Gràcia-Condal}, \citenamefont {Emre}, \citenamefont {Atakol}, \citenamefont
  {Popescu}, \citenamefont {Zhang}, \citenamefont {Long}, \citenamefont
  {Balicas}, \citenamefont {Lluís~Tamarit}, \citenamefont {Planes},
  \citenamefont {Shatruk},\ and\ \citenamefont {Mañosa}}]{romanini19}%
  \BibitemOpen
  \bibfield  {author} {\bibinfo {author} {\bibfnamefont {M.}~\bibnamefont
  {Romanini}}, \bibinfo {author} {\bibfnamefont {Y.}~\bibnamefont {Wang}},
  \bibinfo {author} {\bibfnamefont {K.}~\bibnamefont {Gürpinar}}, \bibinfo
  {author} {\bibfnamefont {G.}~\bibnamefont {Ornelas}}, \bibinfo {author}
  {\bibfnamefont {P.}~\bibnamefont {Lloveras}}, \bibinfo {author}
  {\bibfnamefont {Y.}~\bibnamefont {Zhang}}, \bibinfo {author} {\bibfnamefont
  {W.}~\bibnamefont {Zheng}}, \bibinfo {author} {\bibfnamefont
  {M.}~\bibnamefont {Barrio}}, \bibinfo {author} {\bibfnamefont
  {A.}~\bibnamefont {Aznar}}, \bibinfo {author} {\bibfnamefont
  {A.}~\bibnamefont {Gràcia-Condal}}, \bibinfo {author} {\bibfnamefont
  {B.}~\bibnamefont {Emre}}, \bibinfo {author} {\bibfnamefont {O.}~\bibnamefont
  {Atakol}}, \bibinfo {author} {\bibfnamefont {C.}~\bibnamefont {Popescu}},
  \bibinfo {author} {\bibfnamefont {H.}~\bibnamefont {Zhang}}, \bibinfo
  {author} {\bibfnamefont {Y.}~\bibnamefont {Long}}, \bibinfo {author}
  {\bibfnamefont {L.}~\bibnamefont {Balicas}}, \bibinfo {author} {\bibfnamefont
  {J.}~\bibnamefont {Lluís~Tamarit}}, \bibinfo {author} {\bibfnamefont
  {A.}~\bibnamefont {Planes}}, \bibinfo {author} {\bibfnamefont
  {M.}~\bibnamefont {Shatruk}}, \ and\ \bibinfo {author} {\bibfnamefont
  {L.}~\bibnamefont {Mañosa}},\ }\href {\doibase
  https://doi.org/10.1002/adma.202008076} {\bibfield  {journal} {\bibinfo
  {journal} {Advanced Materials}\ }\textbf {\bibinfo {volume} {33}},\ \bibinfo
  {pages} {2008076} (\bibinfo {year} {2021})}\BibitemShut {NoStop}%
\bibitem [{\citenamefont {Li}\ \emph {et~al.}(2020{\natexlab{a}})\citenamefont
  {Li}, \citenamefont {Li}, \citenamefont {Xu}, \citenamefont {Yang},
  \citenamefont {Xu}, \citenamefont {Jia}, \citenamefont {Li}, \citenamefont
  {He}, \citenamefont {Li},\ and\ \citenamefont {Wang}}]{li20}%
  \BibitemOpen
  \bibfield  {author} {\bibinfo {author} {\bibfnamefont {F.~B.}\ \bibnamefont
  {Li}}, \bibinfo {author} {\bibfnamefont {M.}~\bibnamefont {Li}}, \bibinfo
  {author} {\bibfnamefont {X.}~\bibnamefont {Xu}}, \bibinfo {author}
  {\bibfnamefont {Z.~C.}\ \bibnamefont {Yang}}, \bibinfo {author}
  {\bibfnamefont {H.}~\bibnamefont {Xu}}, \bibinfo {author} {\bibfnamefont
  {C.~K.}\ \bibnamefont {Jia}}, \bibinfo {author} {\bibfnamefont
  {K.}~\bibnamefont {Li}}, \bibinfo {author} {\bibfnamefont {J.}~\bibnamefont
  {He}}, \bibinfo {author} {\bibfnamefont {B.}~\bibnamefont {Li}}, \ and\
  \bibinfo {author} {\bibfnamefont {H.}~\bibnamefont {Wang}},\ }\href {\doibase
  10.1038/s41467-020-18043-1} {\bibfield  {journal} {\bibinfo  {journal}
  {Nature Communications}\ }\textbf {\bibinfo {volume} {11}},\ \bibinfo {pages}
  {4190} (\bibinfo {year} {2020}{\natexlab{a}})}\BibitemShut {NoStop}%
\bibitem [{\citenamefont {Li}\ \emph {et~al.}(2022)\citenamefont {Li},
  \citenamefont {Li}, \citenamefont {Niu},\ and\ \citenamefont {Wang}}]{hui22}%
  \BibitemOpen
  \bibfield  {author} {\bibinfo {author} {\bibfnamefont {F.}~\bibnamefont
  {Li}}, \bibinfo {author} {\bibfnamefont {M.}~\bibnamefont {Li}}, \bibinfo
  {author} {\bibfnamefont {C.}~\bibnamefont {Niu}}, \ and\ \bibinfo {author}
  {\bibfnamefont {H.}~\bibnamefont {Wang}},\ }\href {\doibase
  10.1063/5.0081930} {\bibfield  {journal} {\bibinfo  {journal} {Applied
  Physics Letters}\ }\textbf {\bibinfo {volume} {120}},\ \bibinfo {pages}
  {073902} (\bibinfo {year} {2022})}\BibitemShut {NoStop}%
\bibitem [{\citenamefont {{de Oliveira}}(2023)}]{oliveira23}%
  \BibitemOpen
  \bibfield  {author} {\bibinfo {author} {\bibfnamefont {N.}~\bibnamefont {{de
  Oliveira}}},\ }\href {\doibase https://doi.org/10.1016/j.actamat.2022.118657}
  {\bibfield  {journal} {\bibinfo  {journal} {Acta Materialia}\ }\textbf
  {\bibinfo {volume} {246}},\ \bibinfo {pages} {118657} (\bibinfo {year}
  {2023})}\BibitemShut {NoStop}%
\bibitem [{\citenamefont {Sau}\ \emph {et~al.}(2021{\natexlab{a}})\citenamefont
  {Sau}, \citenamefont {Ikeshoji}, \citenamefont {Takagi}, \citenamefont
  {Orimo}, \citenamefont {Errandonea}, \citenamefont {Chu},\ and\ \citenamefont
  {Cazorla}}]{sau21b}%
  \BibitemOpen
  \bibfield  {author} {\bibinfo {author} {\bibfnamefont {K.}~\bibnamefont
  {Sau}}, \bibinfo {author} {\bibfnamefont {T.}~\bibnamefont {Ikeshoji}},
  \bibinfo {author} {\bibfnamefont {S.}~\bibnamefont {Takagi}}, \bibinfo
  {author} {\bibfnamefont {S.-i.}\ \bibnamefont {Orimo}}, \bibinfo {author}
  {\bibfnamefont {D.}~\bibnamefont {Errandonea}}, \bibinfo {author}
  {\bibfnamefont {D.}~\bibnamefont {Chu}}, \ and\ \bibinfo {author}
  {\bibfnamefont {C.}~\bibnamefont {Cazorla}},\ }\href {\doibase
  10.1038/s41598-021-91123-4} {\bibfield  {journal} {\bibinfo  {journal}
  {Scientific Reports}\ }\textbf {\bibinfo {volume} {11}},\ \bibinfo {pages}
  {11915} (\bibinfo {year} {2021}{\natexlab{a}})}\BibitemShut {NoStop}%
\bibitem [{\citenamefont {Skripov}\ \emph {et~al.}(2015)\citenamefont
  {Skripov}, \citenamefont {Skoryunov}, \citenamefont {Soloninin},
  \citenamefont {Babanova}, \citenamefont {Tang}, \citenamefont {Stavila},\
  and\ \citenamefont {Udovic}}]{skripov15}%
  \BibitemOpen
  \bibfield  {author} {\bibinfo {author} {\bibfnamefont {A.~V.}\ \bibnamefont
  {Skripov}}, \bibinfo {author} {\bibfnamefont {R.~V.}\ \bibnamefont
  {Skoryunov}}, \bibinfo {author} {\bibfnamefont {A.~V.}\ \bibnamefont
  {Soloninin}}, \bibinfo {author} {\bibfnamefont {O.~A.}\ \bibnamefont
  {Babanova}}, \bibinfo {author} {\bibfnamefont {W.~S.}\ \bibnamefont {Tang}},
  \bibinfo {author} {\bibfnamefont {V.}~\bibnamefont {Stavila}}, \ and\
  \bibinfo {author} {\bibfnamefont {T.~J.}\ \bibnamefont {Udovic}},\ }\href
  {\doibase 10.1021/acs.jpcc.5b10055} {\bibfield  {journal} {\bibinfo
  {journal} {The Journal of Physical Chemistry C}\ }\textbf {\bibinfo {volume}
  {119}},\ \bibinfo {pages} {26912} (\bibinfo {year} {2015})}\BibitemShut
  {NoStop}%
\bibitem [{\citenamefont {Mohtadi}\ and\ \citenamefont
  {Orimo}(2016)}]{Mohtadi16}%
  \BibitemOpen
  \bibfield  {author} {\bibinfo {author} {\bibfnamefont {R.}~\bibnamefont
  {Mohtadi}}\ and\ \bibinfo {author} {\bibfnamefont {S.-i.}\ \bibnamefont
  {Orimo}},\ }\href {\doibase 10.1038/natrevmats.2016.91} {\bibfield  {journal}
  {\bibinfo  {journal} {Nature Reviews Materials}\ }\textbf {\bibinfo {volume}
  {2}},\ \bibinfo {pages} {16091} (\bibinfo {year} {2016})}\BibitemShut
  {NoStop}%
\bibitem [{\citenamefont {Dimitrievska}\ \emph {et~al.}(2020)\citenamefont
  {Dimitrievska}, \citenamefont {Wu}, \citenamefont {Stavila}, \citenamefont
  {Babanova}, \citenamefont {Skoryunov}, \citenamefont {Soloninin},
  \citenamefont {Zhou}, \citenamefont {Trump}, \citenamefont {Andersson},
  \citenamefont {Skripov},\ and\ \citenamefont {Udovic}}]{udovic20}%
  \BibitemOpen
  \bibfield  {author} {\bibinfo {author} {\bibfnamefont {M.}~\bibnamefont
  {Dimitrievska}}, \bibinfo {author} {\bibfnamefont {H.}~\bibnamefont {Wu}},
  \bibinfo {author} {\bibfnamefont {V.}~\bibnamefont {Stavila}}, \bibinfo
  {author} {\bibfnamefont {O.~A.}\ \bibnamefont {Babanova}}, \bibinfo {author}
  {\bibfnamefont {R.~V.}\ \bibnamefont {Skoryunov}}, \bibinfo {author}
  {\bibfnamefont {A.~V.}\ \bibnamefont {Soloninin}}, \bibinfo {author}
  {\bibfnamefont {W.}~\bibnamefont {Zhou}}, \bibinfo {author} {\bibfnamefont
  {B.~A.}\ \bibnamefont {Trump}}, \bibinfo {author} {\bibfnamefont {M.~S.}\
  \bibnamefont {Andersson}}, \bibinfo {author} {\bibfnamefont {A.~V.}\
  \bibnamefont {Skripov}}, \ and\ \bibinfo {author} {\bibfnamefont {T.~J.}\
  \bibnamefont {Udovic}},\ }\href {\doibase 10.1021/acs.jpcc.0c05038}
  {\bibfield  {journal} {\bibinfo  {journal} {The Journal of Physical Chemistry
  C}\ }\textbf {\bibinfo {volume} {124}},\ \bibinfo {pages} {17992} (\bibinfo
  {year} {2020})}\BibitemShut {NoStop}%
\bibitem [{\citenamefont {Kim}\ \emph {et~al.}(2019)\citenamefont {Kim},
  \citenamefont {Oguchi}, \citenamefont {Toyama}, \citenamefont {Sato},
  \citenamefont {Takagi}, \citenamefont {Otomo}, \citenamefont {Arunkumar},
  \citenamefont {Kuwata}, \citenamefont {Kawamura},\ and\ \citenamefont
  {Orimo}}]{Kim2019}%
  \BibitemOpen
  \bibfield  {author} {\bibinfo {author} {\bibfnamefont {S.}~\bibnamefont
  {Kim}}, \bibinfo {author} {\bibfnamefont {H.}~\bibnamefont {Oguchi}},
  \bibinfo {author} {\bibfnamefont {N.}~\bibnamefont {Toyama}}, \bibinfo
  {author} {\bibfnamefont {T.}~\bibnamefont {Sato}}, \bibinfo {author}
  {\bibfnamefont {S.}~\bibnamefont {Takagi}}, \bibinfo {author} {\bibfnamefont
  {T.}~\bibnamefont {Otomo}}, \bibinfo {author} {\bibfnamefont
  {D.}~\bibnamefont {Arunkumar}}, \bibinfo {author} {\bibfnamefont
  {N.}~\bibnamefont {Kuwata}}, \bibinfo {author} {\bibfnamefont
  {J.}~\bibnamefont {Kawamura}}, \ and\ \bibinfo {author} {\bibfnamefont
  {S.-i.}\ \bibnamefont {Orimo}},\ }\href {\doibase 10.1038/s41467-019-09061-9}
  {\bibfield  {journal} {\bibinfo  {journal} {Nature Communications}\ }\textbf
  {\bibinfo {volume} {10}},\ \bibinfo {pages} {1081} (\bibinfo {year}
  {2019})}\BibitemShut {NoStop}%
\bibitem [{\citenamefont {Kim}\ \emph {et~al.}(2020)\citenamefont {Kim},
  \citenamefont {Kisu}, \citenamefont {Takagi}, \citenamefont {Oguchi},\ and\
  \citenamefont {Orimo}}]{Kim20204831}%
  \BibitemOpen
  \bibfield  {author} {\bibinfo {author} {\bibfnamefont {S.}~\bibnamefont
  {Kim}}, \bibinfo {author} {\bibfnamefont {K.}~\bibnamefont {Kisu}}, \bibinfo
  {author} {\bibfnamefont {S.}~\bibnamefont {Takagi}}, \bibinfo {author}
  {\bibfnamefont {H.}~\bibnamefont {Oguchi}}, \ and\ \bibinfo {author}
  {\bibfnamefont {S.-I.}\ \bibnamefont {Orimo}},\ }\href {\doibase
  10.1021/acsaem.0c00433} {\bibfield  {journal} {\bibinfo  {journal} {ACS
  Applied Energy Materials}\ }\textbf {\bibinfo {volume} {3}},\ \bibinfo
  {pages} {4831} (\bibinfo {year} {2020})}\BibitemShut {NoStop}%
\bibitem [{\citenamefont {Sau}\ \emph {et~al.}(2021{\natexlab{b}})\citenamefont
  {Sau}, \citenamefont {Ikeshoji}, \citenamefont {Kim}, \citenamefont
  {Takagi},\ and\ \citenamefont {Orimo}}]{Sau20212357}%
  \BibitemOpen
  \bibfield  {author} {\bibinfo {author} {\bibfnamefont {K.}~\bibnamefont
  {Sau}}, \bibinfo {author} {\bibfnamefont {T.}~\bibnamefont {Ikeshoji}},
  \bibinfo {author} {\bibfnamefont {S.}~\bibnamefont {Kim}}, \bibinfo {author}
  {\bibfnamefont {S.}~\bibnamefont {Takagi}}, \ and\ \bibinfo {author}
  {\bibfnamefont {S.~I.}\ \bibnamefont {Orimo}},\ }\href {\doibase
  10.1021/acs.chemmater.0c04473} {\bibfield  {journal} {\bibinfo  {journal}
  {Chemistry of Materials}\ }\textbf {\bibinfo {volume} {33}},\ \bibinfo
  {pages} {2357} (\bibinfo {year} {2021}{\natexlab{b}})}\BibitemShut {NoStop}%
\bibitem [{\citenamefont {Li}\ \emph {et~al.}(2020{\natexlab{b}})\citenamefont
  {Li}, \citenamefont {Dunstan}, \citenamefont {Lou}, \citenamefont {Planes},
  \citenamefont {Ma{\~n}osa}, \citenamefont {Barrio}, \citenamefont {Tamarit},\
  and\ \citenamefont {Lloveras}}]{Li2020reversible}%
  \BibitemOpen
  \bibfield  {author} {\bibinfo {author} {\bibfnamefont {J.}~\bibnamefont
  {Li}}, \bibinfo {author} {\bibfnamefont {D.}~\bibnamefont {Dunstan}},
  \bibinfo {author} {\bibfnamefont {X.}~\bibnamefont {Lou}}, \bibinfo {author}
  {\bibfnamefont {A.}~\bibnamefont {Planes}}, \bibinfo {author} {\bibfnamefont
  {L.}~\bibnamefont {Ma{\~n}osa}}, \bibinfo {author} {\bibfnamefont
  {M.}~\bibnamefont {Barrio}}, \bibinfo {author} {\bibfnamefont {J.-L.}\
  \bibnamefont {Tamarit}}, \ and\ \bibinfo {author} {\bibfnamefont
  {P.}~\bibnamefont {Lloveras}},\ }\href@noop {} {\bibfield  {journal}
  {\bibinfo  {journal} {Journal of Materials Chemistry A}\ }\textbf {\bibinfo
  {volume} {8}},\ \bibinfo {pages} {20354} (\bibinfo {year}
  {2020}{\natexlab{b}})}\BibitemShut {NoStop}%
\bibitem [{\citenamefont {Plimpton}(1995)}]{lammps}%
  \BibitemOpen
  \bibfield  {author} {\bibinfo {author} {\bibfnamefont {S.}~\bibnamefont
  {Plimpton}},\ }\href {\doibase https://doi.org/10.1006/jcph.1995.1039}
  {\bibfield  {journal} {\bibinfo  {journal} {Journal of Computational
  Physics}\ }\textbf {\bibinfo {volume} {117}},\ \bibinfo {pages} {1} (\bibinfo
  {year} {1995})}\BibitemShut {NoStop}%
\bibitem [{\citenamefont {Kresse}\ and\ \citenamefont
  {Hafner}(1993)}]{kresse1993ab}%
  \BibitemOpen
  \bibfield  {author} {\bibinfo {author} {\bibfnamefont {G.}~\bibnamefont
  {Kresse}}\ and\ \bibinfo {author} {\bibfnamefont {J.}~\bibnamefont
  {Hafner}},\ }\href@noop {} {\bibfield  {journal} {\bibinfo  {journal} {Phys.
  Rev. B}\ }\textbf {\bibinfo {volume} {47}},\ \bibinfo {pages} {558} (\bibinfo
  {year} {1993})}\BibitemShut {NoStop}%
\bibitem [{\citenamefont {Perdew}\ \emph {et~al.}(1996)\citenamefont {Perdew},
  \citenamefont {Burke},\ and\ \citenamefont {Ernzerhof}}]{Perdew19963865}%
  \BibitemOpen
  \bibfield  {author} {\bibinfo {author} {\bibfnamefont {J.}~\bibnamefont
  {Perdew}}, \bibinfo {author} {\bibfnamefont {K.}~\bibnamefont {Burke}}, \
  and\ \bibinfo {author} {\bibfnamefont {M.}~\bibnamefont {Ernzerhof}},\ }\href
  {\doibase 10.1103/PhysRevLett.77.3865} {\bibfield  {journal} {\bibinfo
  {journal} {Physical Review Letters}\ }\textbf {\bibinfo {volume} {77}},\
  \bibinfo {pages} {3865} (\bibinfo {year} {1996})}\BibitemShut {NoStop}%
\bibitem [{\citenamefont {Blöchl}(1994)}]{Blochl199417953}%
  \BibitemOpen
  \bibfield  {author} {\bibinfo {author} {\bibfnamefont {P.}~\bibnamefont
  {Blöchl}},\ }\href {\doibase 10.1103/PhysRevB.50.17953} {\bibfield
  {journal} {\bibinfo  {journal} {Physical Review B}\ }\textbf {\bibinfo
  {volume} {50}},\ \bibinfo {pages} {17953} (\bibinfo {year}
  {1994})}\BibitemShut {NoStop}%
\bibitem [{\citenamefont {Sagotra}\ \emph
  {et~al.}(2019{\natexlab{a}})\citenamefont {Sagotra}, \citenamefont {Chu},\
  and\ \citenamefont {Cazorla}}]{Sagotra2019}%
  \BibitemOpen
  \bibfield  {author} {\bibinfo {author} {\bibfnamefont {A.~K.}\ \bibnamefont
  {Sagotra}}, \bibinfo {author} {\bibfnamefont {D.}~\bibnamefont {Chu}}, \ and\
  \bibinfo {author} {\bibfnamefont {C.}~\bibnamefont {Cazorla}},\ }\href
  {\doibase 10.1103/PhysRevMaterials.3.035405} {\bibfield  {journal} {\bibinfo
  {journal} {Phys. Rev. Mater.}\ }\textbf {\bibinfo {volume} {3}},\ \bibinfo
  {pages} {035405} (\bibinfo {year} {2019}{\natexlab{a}})}\BibitemShut
  {NoStop}%
\bibitem [{\citenamefont {Sagotra}\ \emph
  {et~al.}(2019{\natexlab{b}})\citenamefont {Sagotra}, \citenamefont {Chu},\
  and\ \citenamefont {Cazorla}}]{cazorla19b}%
  \BibitemOpen
  \bibfield  {author} {\bibinfo {author} {\bibfnamefont {A.~K.}\ \bibnamefont
  {Sagotra}}, \bibinfo {author} {\bibfnamefont {D.}~\bibnamefont {Chu}}, \ and\
  \bibinfo {author} {\bibfnamefont {C.}~\bibnamefont {Cazorla}},\ }\href
  {\doibase 10.1103/PhysRevMaterials.3.035405} {\bibfield  {journal} {\bibinfo
  {journal} {Physical Review Materials}\ }\textbf {\bibinfo {volume} {3}},\
  \bibinfo {pages} {035405} (\bibinfo {year} {2019}{\natexlab{b}})}\BibitemShut
  {NoStop}%
\bibitem [{\citenamefont {Togo}\ \emph {et~al.}(2010)\citenamefont {Togo},
  \citenamefont {Chaput}, \citenamefont {Tanaka},\ and\ \citenamefont
  {Hug}}]{qha}%
  \BibitemOpen
  \bibfield  {author} {\bibinfo {author} {\bibfnamefont {A.}~\bibnamefont
  {Togo}}, \bibinfo {author} {\bibfnamefont {L.}~\bibnamefont {Chaput}},
  \bibinfo {author} {\bibfnamefont {I.}~\bibnamefont {Tanaka}}, \ and\ \bibinfo
  {author} {\bibfnamefont {G.}~\bibnamefont {Hug}},\ }\href {\doibase
  10.1103/PhysRevB.81.174301} {\bibfield  {journal} {\bibinfo  {journal}
  {Physical Review B}\ }\textbf {\bibinfo {volume} {81}},\ \bibinfo {pages}
  {174301} (\bibinfo {year} {2010})}\BibitemShut {NoStop}%
\bibitem [{\citenamefont {Takagi}\ \emph {et~al.}(2020)\citenamefont {Takagi},
  \citenamefont {Ikeshoji}, \citenamefont {Sato},\ and\ \citenamefont
  {Orimo}}]{takagi20}%
  \BibitemOpen
  \bibfield  {author} {\bibinfo {author} {\bibfnamefont {S.}~\bibnamefont
  {Takagi}}, \bibinfo {author} {\bibfnamefont {T.}~\bibnamefont {Ikeshoji}},
  \bibinfo {author} {\bibfnamefont {T.}~\bibnamefont {Sato}}, \ and\ \bibinfo
  {author} {\bibfnamefont {S.-i.}\ \bibnamefont {Orimo}},\ }\href {\doibase
  10.1063/5.0002992} {\bibfield  {journal} {\bibinfo  {journal} {Applied
  Physics Letters}\ }\textbf {\bibinfo {volume} {116}},\ \bibinfo {pages}
  {173901} (\bibinfo {year} {2020})}\BibitemShut {NoStop}%
\end{thebibliography}%

\section*{Acknowledgements}
K.S. thanks the JSPS International Fellowship. This work was supported by JSPS KAKENHI Grant-in-Aid for Scientific Research on Innovative Areas “Hydrogenomics”, No. JP18H05513 and JSPS Fellowship grant (21F21345). We gratefully acknowledge the Center for Computational Materials Science of Institute for Materials Research, Tohoku University for permitting the use of MASAMUNE- IMR (MAterial science Supercomputing system for Advanced MUltiscale simulations toward Next-generation Institute of Material Research) (project no. 202203-SCKXX-0408). This work was also supported by MINECO Project No.~PID2020-112975GB-I00 (Spain) and DGU Project No.~2021SGR-00343 (Catalonia). C.C. acknowledges financial support from the Spanish Ministry of Science, Innovation and Universities under the ``Ram\'on y Cajal'' fellowship RYC2018-024947-I and the TED2021-130265B-C22 project, and computational support from the Red Espa\~nola de Supercomputaci\'on (RES) under the grants FI-2022-1-0006, FI-2022-2-0003 and FI-2022-3-0014. M.Z. (CSC No. 202008310208) thanks the China Scholarship Council.
\\

\section*{Conflict of Interest}
The authors declare no conflict of interest.
\\

\section*{Data Availability Statement}
The  data  that  support  the  findings  of  this  study  are  available  from  the  corresponding authors upon reasonable request.



\end{document}